\documentclass[%
reprint,
nofootinbib,
mathtools,
amsmath,
amssymb,
prd,
longbibliography,
superscriptaddress]{revtex4-2}

\usepackage{booktabs}
\usepackage{mathrsfs}
\usepackage{graphicx}% Include figure files
\usepackage{dcolumn}% Align table columns on decimal point
\usepackage{bm}% bold math
\usepackage{lipsum}
\usepackage{graphics}
\usepackage{bbm}
\usepackage[dvipsnames]{xcolor}
\usepackage{color}
\usepackage{microtype}
\usepackage{IEEEtrantools}
\usepackage{mathrsfs}
\usepackage{amsthm}
\usepackage{enumitem}
\usepackage[normalem]{ulem}
\usepackage[pdftex,breaklinks,colorlinks,
linkcolor=Blue,
citecolor=teal,
anchorcolor=red,
urlcolor=cyan]{hyperref}
\hypersetup{breaklinks=true}
\usepackage[T1]{fontenc}% For thorn and eth
\usepackage{lmodern}
\usepackage{orcidlink}
\usepackage{empheq} % for multiline boxed equations

\usepackage{url}
\usepackage[mathscr]{euscript} % For \mathscr

\usepackage[latin9]{inputenc}
\usepackage{multirow}
\usepackage{tabularray}

\usepackage{esint}
\usepackage{soul}

\usepackage{amsfonts}
\usepackage{babel}

\usepackage[capitalise]{cleveref}

\usepackage{verbatim}

\graphicspath{{Plots/}}
\font\ec=ecrm0800 at 10pt

\def\pth{\hbox{\ec\char'336}'}

\def\l{\ell}
\def\m{\mathtt{m}}

\def\n{\mathtt{n}}

\def\lTwo{L}
\def\mTwo{M}

\def\scri{\mathscr{I}}
\def\horizon{\mathscr{H}}

\renewcommand{\varsigma}{s}

\begin{document}

\title{Quadratic Quasinormal Mode Dependence on Linear Mode Parity}

\author{Patrick Bourg\orcidlink{0000-0003-0015-0861}}
\affiliation{Institute for Mathematics, Astrophysics and Particle Physics, Radboud University, Heyendaalseweg 135, 6525 AJ Nijmegen, The Netherlands}
\author{Rodrigo Panosso Macedo\orcidlink{0000-0003-2942-5080}}
\affiliation{%
Niels Bohr International Academy, Niels Bohr Institute, Blegdamsvej 17, 2100 Copenhagen, Denmark
}%
\author{Andrew Spiers\orcidlink{0000-0003-0222-7578}}
\affiliation{%
 School of Mathematical Sciences \& School of Physics and Astronomy,
University of Nottingham, University Park, Nottingham, NG7 2RD, United Kingdom
}%
\affiliation{Nottingham Centre of Gravity, University of Nottingham, University Park, Nottingham, NG7 2RD, United Kingdom}
\author{Benjamin Leather\orcidlink{0000-0001-6186-7271}}
\affiliation{%
Max Planck Institute for Gravitational Physics (Albert Einstein Institute), Am M\"uhlenberg 1, Potsdam 14476, Germany
}%

\author{B\'eatrice Bonga\orcidlink{0000-0002-5808-9517}}
\affiliation{Institute for Mathematics, Astrophysics and Particle Physics, Radboud University, Heyendaalseweg 135, 6525 AJ Nijmegen, The Netherlands}

\author{Adam Pound\orcidlink{0000-0001-9446-0638}}
\affiliation{%
 School of Mathematical Sciences and STAG Research Centre, University of Southampton, Southampton, SO17 1BJ, United Kingdom
}

\date{\today}

\begin{abstract}
Quasinormal modes (QNMs) uniquely describe the dominant piece of the gravitational-wave ringdown of postmerger black holes. While the linear QNM regime has been extensively studied, recent work has highlighted the importance of second-perturbative-order, quadratic QNMs (QQNMs) arising from the
nonlinear coupling of linear QNMs. Previous attempts to quantify the magnitude of these QQNMs have shown discrepant results. Using a new hyperboloidal framework, we resolve the discrepancy by showing that the QQNM/QNM ratio is a function not only of the black hole parameters but also of the ratio between even- and odd-parity linear QNMs: the ratio QQNM/QNM depends on what created the ringing black hole, but \emph{only} through this ratio of even- to odd-parity linear perturbations.

\end{abstract}

\maketitle

%%%%%%%%%%%%%%%%%%%%% Introduction %%%%%%%%%%%%%%%%%%%%%%%%
\emph{Introduction}---After the merger of two black holes~(BHs), the distorted remnant BH rings down towards a stationary state through its emission of gravitational waves (GWs). A significant part of the signal associated with this process is well modeled by a superposition of exponentially damped sinusoids, with complex frequencies given by the so-called quasinormal modes (QNMs)~\cite{Kokkotas:1999bd,Nollert:1996rf,Berti:2009kk,Konoplya:2011qq,Barausse:2014tra}. 
For an isolated system within general relativity (GR), these QNM frequencies are uniquely determined by the mass and spin of the final BH. Each frequency $\omega_{\ell \m \n}$ is characterized by three integers: polar and azimuthal indices $(\ell, \m)$ associated with a projection onto spherical harmonics on the celestial sphere, and an overtone index $\n=0,1,\ldots$ that enumerates the frequencies for a given angular mode.

Inspired by the uniqueness of the QNM spectrum, the BH spectroscopy program~\cite{Dreyer:2003bv,Berti:2005ys,Berti:2016lat} aims at extracting multiple QNMs from ringdown signals in order to perform stringent tests of GR, probe the BH geometry, and constrain features of the surrounding environment~\cite{Kokkotas:1999bd,Berti:2009kk,Konoplya:2011qq}. 
Measurements of the dominant QNM, $(\ell,\m,\n)=(2, \pm2, 0)$, in GW signals are well established~\cite{LIGOScientific:2016aoc,LIGOScientific:2020tif,LIGOScientific:2021sio}. The detection of higher overtones and higher angular modes is still under debate~\cite{Isi:2019aib,Capano:2020dix,Capano:2021etf,Cotesta:2022pci,Capano:2022zqm,Forteza:2022tgq,Finch:2022ynt,Abedi:2023kot,Carullo:2023gtf,Baibhav:2023clw,Nee:2023osy,Zhu:2023mzv,Siegel:2023lxl, Gennari:2023gmx}, but future GW detectors such as the Einstein Telescope and LISA are expected to observe these higher modes regularly. Current forecasts predict $20$--$50$ events per year with at least two detectable QNMs for stellar-mass binaries~\cite{Maggiore:2019uih,Cabero:2019zyt} and even $\sim 5$--$8$ QNMs for massive BH binaries with LISA~\cite{Berti:2016lat,Toubiana:2023cwr}.

Historically, the BH spectroscopy program has been entirely based on linear BH perturbation theory (BHPT)~\cite{Kokkotas:1999bd,Berti:2009kk,Konoplya:2011qq,Barausse:2014tra,Dreyer:2003bv,Berti:2005ys}, and forecasts for future QNM detections assume only linear QNM frequencies. However, GR is a nonlinear theory, and recent milestone results have shown that BH spectroscopy must also account for second-order, quadratic perturbations, which can dominate over linear overtones~\cite{London:2014cma,Cheung:2022rbm, Mitman:2022qdl, Zlochower:2003yh,Sberna:2021eui,Baibhav:2023clw,Redondo-Yuste:2023seq,Cheung:2023vki,ma2024excitation}.
In these quadratic perturbations, a new set of characteristic frequencies arises: the so-called quadratic QNMs (QQNMs) $\omega_{\ell_1 \m_1 \n_1\times\ell_2 \m_2 \n_2}=\omega_{\ell_1 \m_1 \n_1}+\omega_{\ell_2 \m_2 \n_2}$, which result from the coupling of two linear QNMs. A recent analysis indicates that the Einstein Telescope and Cosmic Explorer could detect QQNMs in up to a few tens of events per year~\cite{Yi:2024elj}. 
While the predictions for LISA depend sensitively on the astrophysical massive BH formation models, the most optimistic scenario allows for up to ${\cal O}(1000)$ events with detectable QQNMs in LISA's nominal 4-year observation time~\cite{Yi:2024elj}.

Spurred by these developments, there has been a spate of recent work devoted to analysing QQNMs and their impact on BH spectroscopy. Most calculations have been based on extracting modes from fully nonlinear numerical relativity (NR) simulations of BH binary evolutions, but a number of recent calculations have also been performed using second-order BHPT~\cite{ma2024excitation,Zhu:2024rej,Bucciotti:2024zyp}. These calculations have generally focused on a single measure of the significance of QQNMs: the ratio between a given QQNM mode amplitude and the amplitude(s) of the linear parent mode(s) that generate it. Perhaps surprisingly, different analyses have led to conflicting values of the ratio, even in the simplest case of nonrotating BHs. 

As was suggested recently~\cite{ma2024excitation}, we find that these discrepancies are due to the fact that the QQNM ratio depends not only on the BH parameters, but also on the properties of the system that created the ringing BH. Specifically, the QQNM ratio depends on the ratio between even- and odd-parity linear QNMs; this second ratio will depend on the degree to which the progenitor system possessed equatorial (up-down) symmetry. Consequently, we observe that reported QQNM/QNM ratios in the current literature often lack sufficient information to describe the relationship between linear and nonlinear QNMs fully.

As a proof of principle, we define, for the first time, the dependence of the QQNM ratio to the even-to-odd linear parity ratio in the simple case of Schwarzschild spacetime. We discuss how our semi-analytical results generalise to the most generic case in the conclusion. Our calculation utilizes a novel code combining two critical components: a hyperboloidal frequency-domain framework that allows us to directly and accurately compute the physical waveform without requiring regularization~\cite{PanossoMacedo:2014dnr,Ansorg:2016ztf,Ammon:2016fru,PanossoMacedo:2018hab,PanossoMacedo:2019npm,Jaramillo:2020tuu,PanossoMacedo:2023qzp,PanossoMacedo:2022fdi}; and a covariant second-order BHPT formalism~\cite{Spiers:2023mor,mySecond-orderTeuk}. By controlling the geometrical aspects of the problem and using the mode-coupling tools of Ref.~\cite{Spiers:2023mor}, we are able to fine-tune the first-order dynamics to single out any number of linear modes, without contamination from any other aspect of the dynamics (such as late tail decay) and obtain the quadratic contribution from the linear even- and odd-parity sectors semi-analytically. Finally, we compare our results to the current literature and show how our framework can be used to obtain more robust values of the QQNM ratio.

%%%%%%%%%%%%%%%%%%%%% BHPT %%%%%%%%%%%%%%%%%%%%%%%%
\emph{Black hole perturbation theory}---In BHPT we expand the spacetime metric in the form $g_{ab}+\varepsilon h^{(1)}_{ab}+\varepsilon^2 h^{(2)}_{ab}+\ldots$, where $g_{ab}$ is a Kerr metric and $\varepsilon=1$ counts perturbative orders. In vacuum, the perturbations satisfy the Einstein equations 
\begin{equation}\label{EFE}
\varepsilon \delta G_{ab}[h^{(1)}_{cd}]+\varepsilon^2\bigl(\delta G_{ab}[h^{(2)}_{cd}]+\delta^2 G_{ab}[h^{(1)}_{cd}]\bigr)+\ldots=0,
\end{equation}
where $\delta G_{ab}$ is the linearized Einstein tensor and $\delta^2 G_{ab}[h^{(1)}_{cd}]$ is quadratic in $h^{(1)}_{cd}$~\cite{Pound:2021qin}.

At linear order all nontrivial information in $h^{(1)}_{ab}$ is encoded in the linearized Weyl scalar $\Psi^{(1)}_4$, which satisfies the vacuum Teukolsky equation
$ {\mathcal{O}}[\Psi_4^{(1)}]=0$,
where $ {\mathcal{O}}$ is a linear second-order differential operator~\cite{Teukolsky:1972my, Teukolsky:1973ha}. 

We adopt compactified hyperboloidal coordinates $(\tau,\sigma,\theta,\varphi)$~\cite{PanossoMacedo:2019npm,Ansorg:2016ztf}, in which constant-$\tau$ slices connect the future horizon $\horizon^+$ (at compactified radial coordinate $\sigma = 1$) to future null infinity $\scri^+$ (at $\sigma=0$). Using the hyperboloidal time $\tau$, we introduce the frequency-domain field ${\psi}_{4}^{(1)}$ via a Laplace (or Fourier) transform,
\begin{equation}
	\psi_{4}^{(1)}(\sigma,\theta,\varphi; \varsigma) = \int_0^\infty {\Psi}_{4}^{(1)}(\tau,\sigma,\theta,\varphi) e^{- \varsigma \tau} d \tau.
\end{equation}
The complex Laplace parameter $\varsigma$ is related to the usual complex frequency $\omega$ by $\varsigma = -i \omega $.
We next separate the Teukolsky equation into an angular and radial part by decomposing $\psi_4^{(1)}$ into spin-weighted spheroidal harmonics~\cite{Teukolsky:1972my, Teukolsky:1973ha,Leaver:1986gd},
\begin{equation}
	\psi_4^{(1)} = \mathcal{Z}(\sigma) \sum_{\l\m} \tilde{\psi}_{\l\m}^{(1)}(\sigma;s) {}_{-2} S_{\l\m}(\theta, \varphi;\varsigma).
\end{equation}
Here $\mathcal{Z}(\sigma)$ serves to factor out the dominant behavior near $\horizon^+$ and $\scri^+$; a linear vacuum perturbation $\Psi^{(1)}_4$ scales quadratically with distance from the horizon near $\horizon^+$ and decays inversely with distance toward $\scri^+$~\cite{
PanossoMacedo:2019npm,
Pound:2021qin}, motivating us to choose $\mathcal{Z}(\sigma) \propto \sigma (1-\sigma)^2$. Because we use hyperboloidal slices~\cite{Zenginoglu:2011jz,PanossoMacedo:2019npm,PanossoMacedo:2023qzp}, the modes $\tilde{\psi}_{\l\m}^{(1)}$ are smooth on the entire domain, $\sigma \in [0,1]$, and because we factor out $\mathcal{Z}$ 
we can directly compute the waveform from $\tilde{\psi}_{\l\m}^{(1)}$ at $\sigma=0$.

The Laplace transform and spheroidal-harmonic expansion leave us with a radial Teukolsky equation, 
$ \mathcal{D}_{\l \m} [\tilde{\psi}_{\l \m}^{(1)}(\sigma;\varsigma)] = 0$,
where $\mathcal{D}_{\l\m}$ is a linear second-order radial operator. In the hyperboloidal setup, linear QNMs are the solutions to this equation that satisfy regularity conditions at both ends ($\sigma=0$ and 1). Such solutions only exist for the countable set of QNM frequencies $\varsigma_{\l\m\n}$, and we denote them $\tilde{\psi}_{\l\m\n}^{(1)}(\sigma) := \tilde{\psi}_{\l\m}^{(1)}(\sigma; \varsigma_{\l\m\n})$.

Given a set of QNM solutions\footnote{We absorb all the higher-order homogeneous QNM perturbations into our definition of $\tilde\psi^{(1)}_{\l\m\n}$ without loss of generality.}, $\tilde\psi^{(1)}_{\l\m\n}$, the inverse Laplace transform yields a time-domain solution of the form \cite{Ansorg:2016ztf}
\begin{equation} \label{eqn:Psi4QNMDecomposition}
\Psi_4^{(1)} = \mathcal{Z}(\sigma) \sum_{\l\m\n} A_{\l\m\n} \tilde{\psi}_{\l\m\n}^{(1)}(\sigma) e^{\varsigma_{\l\m\n} \tau} {}_{-2} S_{\l\m}(\theta,\varphi; s_{\l\m\n}).
\end{equation}
Since QNM solutions are only defined up to an overall constant factor, the $A_{\ell m n}$ are arbitrary (complex) excitation coefficients, and we set $\tilde{\psi}_{\l \m \n}^{(1)} = 1$ at $\scri^+$. In Eq.~\eqref{eqn:Psi4QNMDecomposition}, we neglect the initial prompt response and the late-time tail contributions that generically arise~\cite{Nollert:1996rf}. In fact, the dynamics given by \eqref{eqn:Psi4QNMDecomposition} can be made {\em exact} by fine-tuning the initial data to excite only individual QNMs. Thus, using the hyperboloidal formalism~\cite{Ansorg:2016ztf,PanossoMacedo:2023qzp}, we fix a pure QNM at linear order, which provides us with a clean scenario to study QNM coupling at second order.

\emph{Mirror modes and parity}---For any given oscillatory frequency and decay rate within the BH QNM spectrum, there exist two independent eigenvalues and eigenfunctions.
In the Kerr spacetime, this feature manifests itself via the symmetry between the $+\m$ and $-\m$ QNM frequencies~\cite{Leaver:1986gd, Dhani:2020nik} $(\m \neq 0)$
\begin{align}\label{eq:mirror-frequency}
	\varsigma_{\l-\m n}=\varsigma_{\l\m n}^\star,
\end{align}
which motivates the definition of \textit{regular} and \textit{mirror} QNMs, depending, respectively, on whether ${\rm Im}(\varsigma_{\l\m n}) < 0$ or ${\rm Im}(\varsigma_{\l\m n}) > 0$. For $\m=0$, the two independent eigenvalues correspond to the pair of complex conjugate numbers $s_{\l0n,+} = s_{\l 0n}$ and $s_{\l 0n,-} = s^\star_{\l0n}$. In Schwarzschild, the $\m$-dependence on the frequencies degenerate, and the previous relation is valid for all $\m$.
Additionally, our choice of normalization implies that the associated eigenfunctions are related by $\tilde{\psi}_{\l-\m\n} =\tilde{\psi}_{\l\m\n}^\star$ if $\m\neq 0$, and $\tilde{\psi}_{\l 0 \n, -}= \tilde{\psi}^\star_{\l 0 \n, +}$ for $\m=0$.

Most QNM analyses have exclusively considered the regular QNMs. In~\cite{Dhani:2020nik}, the importance of including mirror QNMs was demonstrated in linear QNM analyses to reduce systematic uncertainties. In this work, we find that mirror modes also play a crucial role in QQNM analysis. 

The ratio between regular and mirror modes, $A_{\l-\m\n}/A_{\l\m\n}$, is directly related to the ratio between even- and odd-parity contributions to the GW. At $\scri^+$, the GW (or more strictly, the shear~\cite{Madler:2016xju}) can be naturally decomposed into even-parity ($Y_{AB}$) and odd-parity ($X_{AB}$) tensor harmonics~\cite{Martel:2005ir,Spiers:2023mor},
\begin{equation}
	h_{AB}^{(1)} = r^2\mathcal{Z}(\sigma)\sum_{\l\m\n} \left( C^+_{\l\m\n} Y_{AB}^{\l\m} + C^-_{\l\m\n} X_{AB}^{\l\m} \right) e^{ \varsigma_{\l\m\n} \tau},
\end{equation}
where it is understood that this applies over the sphere with coordinates $\theta^A=(\theta,\varphi)$, in the limit $r\to\infty$ ($\sigma\to0$). The factor of $r^2$ corresponds to the natural scaling of angular components, $C^\pm_{\l\m\n}$ are constant (complex) amplitudes, and $\tau=u$ is the usual outgoing null coordinate.

For comparison, $\Psi_4^{(1)}$ at $\scri^+$ can be decomposed into the closely related spin-weight $-2$ spherical harmonics ${}_{-2}Y_{\l\m}$ (requiring a projection from spheroidal harmonics, except in the Schwarzschild case where ${}_{-2}S_{\l\m}$ reduces to ${}_{-2}Y_{\l\m}$). To relate the amplitudes $A_{\l\m\n}$ to $C_{\l\m\n}^\pm$, we use the relations between harmonics in Ref.~\cite{Spiers:2023mor} together with the fact that $\lim_{r\rightarrow \infty} \Psi_4^{(1)} = -\frac{1}{2}\lim_{r\rightarrow \infty} \ddot{h}_{\bar{m} \bar{m}}$, where $\bar m^A=\frac{1}{\sqrt{2}r}(1,i\csc\theta)$. A short calculation reveals
\begin{align} \label{eqn:AtoC1}
	A_{\l\m\n} &= -\frac{\varsigma_{\l\m\n}^2}{4} \lambda_{\l,2} \left( C^+_{\l\m\n} - i C^-_{\l\m\n} \right), \\
	A_{\l -\m\n}^\star &= -(-1)^\m \frac{\varsigma_{\l\m\n}^2}{4} \lambda_{\l,2} \left( C^+_{\l\m\n} + i C^-_{\l\m\n} \right), \label{eqn:AtoC2}
\end{align}
where $\lambda_{\l,2} = \sqrt{(\l+2)! / (\l-2)!}$.
Equation~\eqref{eqn:AtoC2} follows from the mirror relation \eqref{eq:mirror-frequency} and the fact that the 4D metric perturbation is real, i.e., $C^\pm_{\l -\m\n} = (-1)^m (C^\pm_{\l\m\n})^\star$.

Equations~\eqref{eqn:AtoC1} and \eqref{eqn:AtoC2} imply that the complex ratio $A_{\l-\m\n}^\star/A_{\l\m\n}$ is a simple function of $C_{\l\m\n}^-/C_{\l\m\n}^+$. We emphasize that while it is impossible to omit $\m<0$ modes from $h^{(1)}_{AB}$ (as they are required for $h^{(1)}_{AB}$ to be real valued), it is possible to omit $\m<0$ modes in $\Psi^{(1)}_4$; this corresponds to the particular ratio $C^+_{\l\m\n}=-iC_{\l\m\n}^-$.

%%%%%%%%%%%%%%%%%%%%% QQNMs: %%%%%%%%%%%%%%%%%%%%%%%%
\emph{Quadratic QNMs}---Like at first order, the second-order contribution to the GW is fully encoded in a linear Weyl scalar $\Psi_{4{\rm L}}^{(2)}$ constructed from $h^{(2)}_{ab}$~\cite{Campanelli:1998jv,mySecond-orderTeuk}. $\Psi_{4{\rm L}}^{(2)}$ satisfies a reduced second-order Teukolsky equation derived from the second-order terms in the Einstein equation~\eqref{EFE}~\cite{Green:2019nam,mySecond-orderTeuk}:
\begin{align}\label{eq:teuk-2nd-order}
	{\mathcal{O}}[\Psi_{4{\rm L}}^{(2)}]=- {\mathcal{S}}[\delta^2G_{ab}[h^{(1)}_{cd}]] := \tilde{\mathcal{S}},
\end{align}
where $\mathcal{S}$ is a linear second-order differential operator~\cite{Pound:2021qin}. 

The source in \cref{eq:teuk-2nd-order} depends quadratically on the first-order metric perturbation; schematically,
\begin{align} \label{eqn:2ndOrderSource_schematicForm}
	\tilde{\mathcal{S}} \sim \nabla\nabla( \nabla h^{(1)}_{ab})^2 \; .
\end{align}
In Schwarzschild~\cite{Spiers:2023mor}, the $\l\m$ modes of $\tilde{\mathcal{S}}$ are readily computed from an arbitrary set of first-order $\l\m$ modes using the \textit{Mathematica} package PerturbationEquations~\cite{warburton2023perturbationequations}. 

We obtain the $\l\m$ modes of $h^{(1)}_{ab}$ using a standard metric reconstruction procedure~\cite{Pound:2021qin} in the outgoing radiation gauge (ORG), in which $h^{(1)}_{ab}$ is computed from a Hertz potential $\Phi^{\rm ORG}$ satisfying the inversion relation
\begin{align}\label{eq:radialInversion}
\Psi_4^{(1)} &= \frac{1}{4} \pth^4 \bar{\Phi}^{\rm ORG},
\end{align}
where $\pth$ is a derivative along ingoing principal null rays~\cite{Pound:2021qin}. In terms of this Hertz potential,
\begin{align}\label{eq:MetricRecons}
h_{ab}^{(1)} &= 2 \mathrm{Re}( {\mathcal{S}}^\dagger \Phi^{\rm ORG})_{ab},
\end{align}
where $ {\mathcal{S}}^\dagger$ denotes the adjoint of $ {\mathcal{S}}$. 

Note that $h^{(1)}_{ab}$ depends on both $\Phi^{\rm ORG}$ and $\bar\Phi^{\rm ORG}$ and therefore on both $\Psi_4$ and $\bar\Psi_4$ via \cref{eq:radialInversion}. Consequently, each QQNM amplitude depends on regular and mirror linear QNM amplitudes. To elucidate this, we consider $\Psi_4^{(1)}$ being composed of a single regular and mirror mode
\begin{align} \label{eqn:Ansatz}
	\Psi^{(1)}_4 = \mathcal{Z}(\sigma) \Bigl\{A_{\l \m \n} \tilde{\psi}_{\l \m \n}^{(1)}(\sigma) e^{\varsigma_{\l \m \n} \tau} {}_{-2} Y_{\l \m}(\theta, \phi)  \nonumber \\
	+ A_{\l -\m \n} \tilde{\psi}_{\l -\m \n}^{(1)}(\sigma)e^{\varsigma_{\l -\m \n} \tau} {}_{-2} Y_{\l -\m} (\theta,\phi) \Bigr\}.
\end{align}

This choice reinforces the argument introduced around Eq.~\eqref{eq:mirror-frequency}: There exist only two independent QNM frequencies, related by complex conjugation, associated to a given oscillation period and decay rate
\footnote{Due to the degeneracy in the Schwarzschild limit, these two QNM frequencies degenerate with the index $\m$. Thus, for a given $\l$ mode and overtone $\n$, the most generic form of ansatz (13) in Schwarzschild would actually include a sum over all $\m$. However, we restrict to Eq. (13) as it already includes all the essential features (see footnote 3), and in order to keep in line with the expected expression in Kerr.}.
Though an extended ansatz including further modes is possible, it would only trigger distinct sectors in the nonlinear coupling, associated with different physical timescales, which in turn can be studied independently.

Equations~\eqref{eqn:2ndOrderSource_schematicForm}--\eqref{eq:MetricRecons} show that the second-order source $\tilde{\mathcal{S}}$ depends quadratically on $\Psi^{(1)}_4$ and $\bar\Psi^{(1)}_4$.
Via our ansatz \eqref{eqn:Ansatz} and~\cref{eq:mirror-frequency} and the coupling between spherical harmonics, the ensuing second-order source will be made of several modes $(\lTwo,\mTwo)$, where $\lTwo\leq 2 \l$ and $\mTwo=-2\m,0,2\m$. For simplicity, we will only focus on the ``top'' case where, from now on, $\lTwo=2\l$. Setting $\mTwo=2\m=2\l=\lTwo$, $\tilde{\mathcal{S}}$ is hence composed of three distinct terms ~\cite{Okuzumi:2008ej,Lagos:2022otp},
\begin{align} \label{eqn:2ndOrderSourceAnsatz}
	\tilde{\mathcal{S}} = \mathcal{Z}(\sigma) \Bigl\{&\tilde{S}^{\lTwo \mTwo}(\sigma) e^{2 \varsigma_{\l \m \n} \tau} {}_{-2} Y_{\lTwo \mTwo}(\theta, \phi) \nonumber\\
	&+ \tilde{S}^{\lTwo 0}(\sigma) e^{2 {\rm Re}(\varsigma_{\l \m \n}) \tau} {}_{-2} Y_{\lTwo 0}(\theta, \phi) \nonumber \\
	&+ \tilde{S}^{\lTwo -\mTwo}(\sigma) e^{2 \varsigma_{\l -\m \n} \tau} {}_{-2} Y_{\lTwo -\mTwo}(\theta, \phi) \Bigr\}.
\end{align}

%%%%%%%%%%%%%%%%%%%%% Figure %%%%%%%%%%%%%%%%%%%%%%%%
\begin{figure*}[ht]
	\centering
	\includegraphics[width= \textwidth]{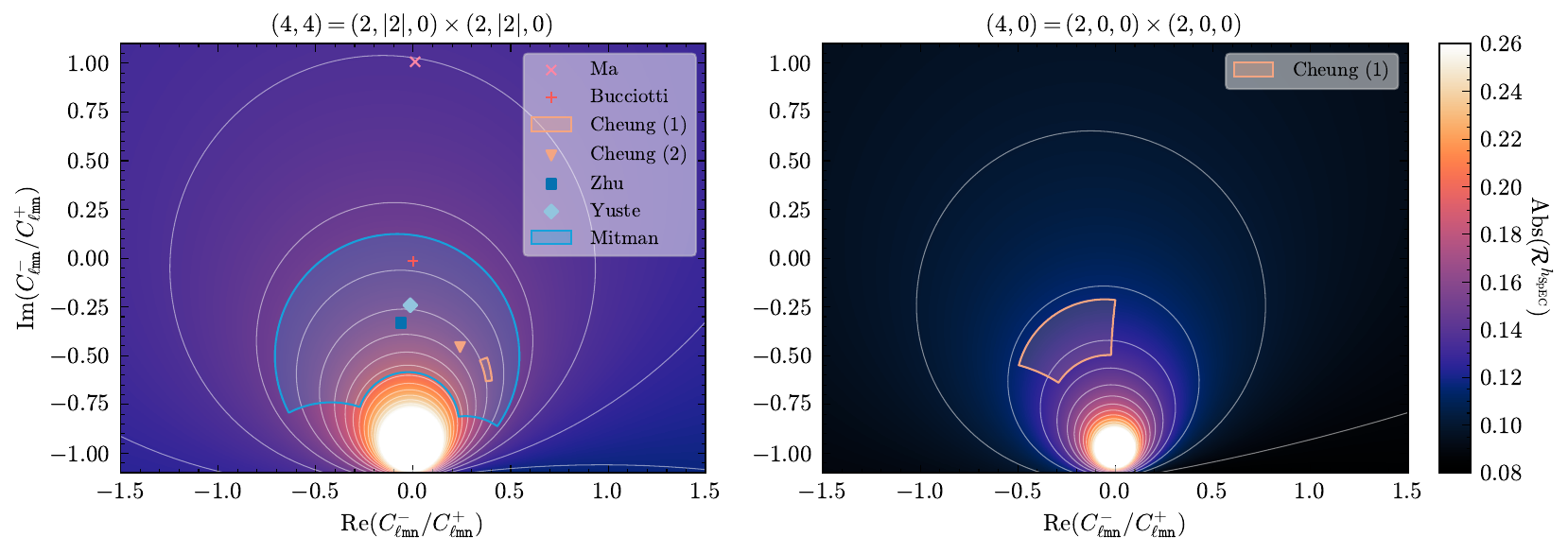}
	\caption{Contour plot of the ratio $\left(\mathcal{R}^{h_{\rm{SpEC}}} \right)^{\lTwo \mTwo}_{(\l \m \n) \times (\l \m \n)}$  as a function of the odd-to-even complex amplitude ratio $C^-_{\l\m\n} / C^+_{\l\m\n}$, with the solid lines indicating values of constant amplitude of the QQNM ratio. {\em Left panel:} Quadratic mode $(4, 4) = (2, |2|, 0) \times (2, |2|, 0)$ with a comparison against prior results employing second-order BHPT:
    Redondo-Yuste \emph{et al.}~\cite{Redondo-Yuste:2023seq};
    Ma \emph{et al.}~\cite{ma2024excitation};
    Bucciotti \emph{et al.}~\cite{Bucciotti:2024zyp};
    Zhu \emph{et al.}~\cite{Zhu:2024rej}.
    We also show results from NR:
    Cheung \emph{et al.}~\cite{Cheung:2022rbm};
    Cheung \emph{et al.}~\cite{Cheung:2023vki};
    Mitman \emph{et al} \cite{Mitman:2022qdl} (which reports a range of values for the magnitude and phase of the QQNM ratio).
	For systems with up-down symmetry, such as those in the NR simulations, all points should lie at the origin $C^-_{\l\m\n}=0$. However, note the NR results are for spinning BHs, while our method of placing them on this plot assumes a nonspinning BH. 
	{\em Right panel:} Quadratic mode $(4, 0) = (2, 0, 0) \times (2, 0, 0)$, which allows a comparison against NR results for head-on collisions yielding a final nonspinning BH.
}
	\label{fig:qqnm_ratio_440}
\end{figure*}

Given the source~\eqref{eqn:2ndOrderSourceAnsatz}, we solve Eq.~\eqref{eq:teuk-2nd-order} following the same procedure as at first order, performing a Laplace transform in hyperboloidal time, decomposing into $\l\m$ modes, and solving the resulting radial equations. The time-domain solution is obtained by applying an inverse Laplace transform, which contains the same modes as the source~\eqref{eqn:2ndOrderSourceAnsatz} (in addition to tails and other contributions that we neglect). In particular, it is composed of $(\lTwo,\mTwo),(\lTwo,0),(\lTwo,-\mTwo)$ spherical modes, with associated frequencies $2 \varsigma_{\l\m\n}$, $2 {\rm Re}(\varsigma_{\l\m\n})$ and $2 \varsigma_{\l-\m\n}$. Each of these QQNMs depends on the first-order excitation coefficients in the following way\footnote{Adding further modes to Ansatz \eqref{eqn:Ansatz} would not affect the functional form of Eqs.~\eqref{eqn:Psi42ndOrder++} and \eqref{eqn:Psi42ndOrder--}, but it would alter Eq.~\eqref{eqn:Psi42ndOrder+-}.}:
\begingroup%
\allowdisplaybreaks
\begin{align}
\label{eqn:Psi42ndOrder++}
	\left(\Psi_{4{\rm L}}^{(2)} \right)^{\lTwo \mTwo} &= \mathcal{Z}(\sigma) \left(a^{\lTwo \mTwo}  (\sigma) A_{\l \m \n}^2 + b^{\lTwo \mTwo}  (\sigma) A_{\l \m \n} A_{\l -\m \n}^\star \right), \\
	\left(\Psi_{4{\rm L}}^{(2)} \right)^{\lTwo 0} &= \mathcal{Z}(\sigma) \left(c^{\lTwo 0}  (\sigma) A_{\l \m \n} A_{\l \m \n}^\star + d^{\lTwo 0} (\sigma) A_{\l \m \n} A_{\l -\m \n}^\star \right. \nonumber \\*
	&\left. \quad + \left(c^{\lTwo 0}  (\sigma)\right)^\star A_{\l -\m \n} A_{\l -\m \n}^\star \right), \label{eqn:Psi42ndOrder+-} \\
	\left(\Psi_{4{\rm L}}^{(2)} \right)^{\lTwo -\mTwo} &= \mathcal{Z}(\sigma) \left(\left(a^{\lTwo \mTwo}  (\sigma) \right)^\star A_{\l -\m \n}^2 \right. \nonumber \\
	& \left. \quad + \left(b^{\lTwo \mTwo}  (\sigma)\right)^\star A_{\l -\m \n} A_{\l \m \n}^\star \right). \label{eqn:Psi42ndOrder--}
\end{align}
\endgroup
We provide the coefficients ($a^{\lTwo \mTwo}$ etc.), evaluated at $\scri^+$, in the Supplemental Material.

Note that one can uniquely determine the two excitation amplitudes of the first-order data, $A_{\l \m \n}$ and $A_{\l -\m \n}$, up to an overall sign, from the excitations at second order, by solving the system \eqref{eqn:Psi42ndOrder++}-\eqref{eqn:Psi42ndOrder--} for $A_{\l \m \n}$ and $A_{\l -\m \n}$ at a radial point $\sigma = \sigma_0$, for example at $\scri^+$, $\sigma_0=0$. Given the relations \eqref{eqn:AtoC1}--\eqref{eqn:AtoC2}, this means that from the QQNMs one can uniquely determine, up to an overall sign (which corresponds to a phase difference of $+\pi$), the contribution of the even and odd sectors of the first-order QNMs.
The residual sign ambiguity is due to the fact that the second-order source, and therefore the QQNMs, are invariant under a sign change $\Psi_4^{(1)} \to -\Psi_4^{(1)}$; see again Eq.~\eqref{eqn:2ndOrderSource_schematicForm}.

%%%%%%%%%%%%%%%%%%%%% Results %%%%%%%%%%%%%%%%%%%%%%%%
\emph{Results}---We specialize to the scenario described above, with first-order pure QNM data consisting of a single regular and mirror mode, as in~\eqref{eqn:Ansatz}. The ensuing nonzero pieces of the second-order Weyl scalar are of the form \eqref{eqn:Psi42ndOrder++}--\eqref{eqn:Psi42ndOrder--}. For concreteness, we display results for the QQNM mode with frequency $2\varsigma_{\l \m \n}$, which corresponds to the expression in Eq.~\eqref{eqn:Psi42ndOrder++}.

Typically, one is not directly interested in the value of $\left(\Psi_{4{\rm L}}^{(2)} \right)^{\lTwo \mTwo} $ at $\scri^+$, but in how it compares to the first-order perturbation. Since $\Psi_{4{\rm L}}^{(2)} \sim \bigl(\Psi_4^{(1)} \bigr)^2$, we might consider the following ratio:
\begin{equation} \label{eqn:QQNMRatio}
    \left(\mathcal{R}^{\Psi_4} \right)^{\lTwo \mTwo}  := \frac{\left(r \Psi_{4{\rm L}}^{(2)} \right)^{\lTwo \mTwo} }{ \left[\left(r \Psi_4^{(1)} \right)_{\l \m \n} \right]^2}.
\end{equation}
where the above is understood to be evaluated at $\scri^+$.
However, for comparison with the literature, it is more convenient to define the analogous ratio from the strain $h$,
\begin{equation} \label{eqn:ratioSpEC}
        \left(\mathcal{R}^{h_{\rm{SpEC}}} \right)^{\lTwo \mTwo} = -\frac{\varsigma_{\l\m\n}^2}{8} \left(\mathcal{R}^{\Psi_4} \right)^{\lTwo \mTwo} .
\end{equation}

Note that, apart from numerical factors, these ratios only depend on the ratio between mirror and regular mode amplitudes, $A_{\l-\m\n}^\star/A_{\l\m\n}$. For example, $\left(\mathcal{R}^{\Psi_4} \right)^{\lTwo \mTwo}/8 = a^{\lTwo \mTwo} (0) + b^{\lTwo \mTwo} (0) A_{\l -\m \n}^\star/A_{\l \m \n}$. Using Eqs.~\eqref{eqn:AtoC1} and \eqref{eqn:AtoC2}, we can alternatively express $\left(\mathcal{R}^{\Psi_4} \right)^{\lTwo \mTwo} $ in terms of the ratio of odd to even amplitudes, $C^-_{\l\m\n} / C^+_{\l\m\n}$. In Kerr spacetime $\left(\mathcal{R}^{\Psi_4}\right)^{\lTwo \mTwo}$ is also a function of the BH parameters and $C^-_{\l\m\n} / C^+_{\l\m\n}$. This can be derived by inputting Eq.~\eqref{eq:MetricRecons} into the expression for the source, $\tilde{\mathcal{S}}$, available in Ref.~\cite{2nd-order-notebook}. This shows that the source consists of terms $\propto \bar\Phi^2$ and $\propto \Phi\bar\Phi$. Hence, using Eqs.~\eqref{eq:radialInversion},~\eqref{eqn:Ansatz}, and~\eqref{eq:mirror-frequency}, similar relations as 
Eqs.~\eqref{eqn:Psi42ndOrder++} to~\eqref{eqn:Psi42ndOrder--} hold in Kerr (up to angular mode mixing) and
$\left(\mathcal{R}^{\Psi_4}\right)^{\lTwo \mTwo}$ is a function of $C^-_{\l\m\n} / C^+_{\l\m\n}$ and the BH parameters.

In Fig.~\ref{fig:qqnm_ratio_440}, we show a contour plot of $\left(\mathcal{R}^{h_{\rm{SpEC}}} \right)^{\lTwo \mTwo} $ as a function of $C^-_{\l \m \n} / C^+_{\l \m \n}$ for the case $(\l,|\m|,\n) = (2,2,0)$ (left plot) and $(\l,\m,\n) = (2,0,0)$ (right plot). Several papers have provided numerical calculations of this ratio \cite{Perrone:2023jzq,Nakano:2007cj,ma2024excitation,Bucciotti:2024zyp,Cheung:2022rbm,Cheung:2023vki,Zhu:2024rej,Mitman:2022qdl,Redondo-Yuste:2023seq}, a selection of which we display in the figure; we relegate to the Supplemental Material how these data points were added to the figure. We highlight the significant difference between the results reported by Ma \emph{et al.}.~\cite{ma2024excitation} (pink cross) and those obtained from NR data (in blue and orange) or from the BHPT result reported by Bucciotti \emph{et al.}~\cite{Bucciotti:2024zyp} (light red plus). Ma \emph{et al.} report a ratio $\approx 0.137e^{-0.083i}$. Their computation of this result assumed that $\Psi_4^{(1)}$ is composed of a single regular frequency, $\varsigma_{\l \m \n}$, with $A_{\l -\m \n} = 0$. Equivalently, they assume $C^-_{\l \m \n} = i C^+_{\l \m \n}$. We find that we exactly recover their result for $\left(\mathcal{R}^{h_{\rm{SpEC}}} \right)^{\lTwo \mTwo} $ in our framework, given their specific ratio $C^-_{\l \m \n}/C^+_{\l \m \n}$.

In contrast, all other data points indicate larger magnitudes for the ratio, typically $\approx 0.15$--$0.20$. This places them much closer to the origin of the figure, where $C^-_{\l\m\n}=0$.
For the NR data we consider, the simulated binary systems possess up-down symmetry; hence, the odd-parity modes should identically vanish for even values of $\l+\m$. Therefore, data for these systems should lie precisely at $C^-_{22n} = C^-_{20n} = 0$. For the case $(2,|2|,0)\times (2,|2|,0)\rightarrow (4,4)$, the reported NR values might appear relatively far away from this point. However, NR simulations of binary inspirals yield spinning remnant BHs, while we use a Schwarzchild BH in computing the relationship between $\left(\mathcal{R}^{h_{\rm{SpEC}}} \right)^{\lTwo \mTwo}_{(\l \m \n) \times (\l \m \n)}$ and $C^-_{\l\m\n}/C^+_{\l\m\n}$. This likely explains much of the deviation of the NR data away from the origin in the left panel of the figure.

To eliminate this uncertainty due to spin, in the right panel of Fig.~\ref{fig:qqnm_ratio_440} we show results for the mode coupling $(2,0,0)\times (2,0,0)\rightarrow (4,0)$. The NR data we display in this panel come from simulations of head-on collisions~\cite{Cheung:2022rbm}, for which the remnant BHs are nonspinning. However, the systems remain up-down symmetric, implying the NR data should lie at $C^-_{20n} = 0$. This prediction is nearly within the lower bound for the error bar reported by the authors. Since these error bars mainly estimate contributions from the fitting, we expect the theoretical prediction $C^-_{20n} = 0$ to lie within the error estimates, once all systematic errors are accounted for. However, the nonzero mean deviation from the origin suggests that the QQNM ratio could be more accurately extracted from NR data if the constraint $C^-_{20n}=0$ were enforced in the fitting procedure.

%%%%%%%%%%%%%%%%%%%%% Discussion %%%%%%%%%%%%%%%%%%%%%%%%
\emph{Discussion}---Precision BH spectroscopy is expected to be a pillar of future GW astronomy, enabling stringent tests of GR and of whether the massive objects in galactic centers are described by the Kerr spacetime. This program is now widely expected to require calculations of QQNMs. In this Letter, we have shown how some disagreements in recent calculations of QQNM amplitudes can be reconciled: both even- and odd-parity linear QNMs (or equivalently, regular and mirror modes) contribute to the same QQNM, and at least some of the discrepancies in the literature are due to differences in the relative excitations of the even and odd sectors. Most importantly, we have shown that in vacuum GR this is the unique way in which the QQNM amplitudes depend on the system that formed the BH (beyond the fact that the BH mass and spin also depend on the progenitor system). Our results therefore suggest that measurements of QQNMs can be used to extract how much the first-order even and odd sectors have been excited, providing a unique route to determining, for example, the breaking of isospectrality in beyond-GR theories or due to environmental effects.

Given our results, an important task for future work will be to explore how the ratio $C^-_{\l\m\n}/C^+_{\l\m\n}$ depends on the details of the binary that formed the final BH. This would, in principle, make it simple to assess how the QQNM ratio depends on the BH's formation. Our results also highlight the importance of enforcing system symmetries (e.g., up-down symmetry) when fitting NR ringdowns.

The results presented here were restricted to Schwarzschild, but our main conclusion and our computational framework generalise readily to Kerr~\cite{PanossoMacedo:2019npm,mySecond-orderTeuk,Spiers:2024src}. A companion paper will provide details of the framework and will be released with a complete code that can easily handle any number of first-order QNMs. The framework may also be applied to study GW memory~\cite{Mitman:2024uss} and quadratic couplings involving GW tails~\cite{Cardoso:2024jme}.

%%%%%%%%%%%%%%%%% Notes added and conclusion %%%%%%%%%%%%%%%%%%%%
\emph{Note added}---Recently, \cite{Bucciotti:2024zyp} appeared, which also emphasised the importance of the mirror modes in QQNM analysis. Their results are compatible with the results in this Letter, but they make specific choices of even and odd parity ratios rather than characterizing the full dependence.

\emph{Acknowledgments}---We gratefully acknowledge helpful discussions with Sizheng Ma, Huan Yang, and Neev Khera. A.S. would like to thank Laura Sberna and Stephen Green for their helpful discussions. B.L. would like to thank Sebastian V\"olkel and Hector Okada da Silva for their helpful discussions. R.P.M. thanks Jaime Redondo-Yuste and Mark Cheung for valuable discussions. P.B. and B.B. acknowledge the support of the Dutch Research Council (NWO) (project name: Resonating with the new gravitational-wave era, Project No. OCENW.M.21.119). R.P.M. acknowledges support from the Villum Investigator program supported by the VILLUM Foundation (Grant No. VIL37766) and the DNRF Chair program (Grant No. DNRF162) by the Danish National Research Foundation and the European Union's Horizon 2020 research and innovation programme under the Marie Sklodowska-Curie Grant Agreement No. 101131233.
A.S. acknowledges support from the STFC Consolidated Grant No. ST/V005596/1. AP acknowledges the support of a Royal Society University Research Fellowship and a UKRI Frontier Research Grant (as selected by the ERC) under the Horizon Europe Guarantee scheme [Grant No. EP/Y008251/1].

\bibliography{references}

%apsrev4-2.bst 2019-01-14 (MD) hand-edited version of apsrev4-1.bst
%Control: key (0)
%Control: author (8) initials jnrlst
%Control: editor formatted (1) identically to author
%Control: production of article title (0) allowed
%Control: page (0) single
%Control: year (1) truncated
%Control: production of eprint (0) enabled
\begin{thebibliography}{68}%
\makeatletter
\providecommand \@ifxundefined [1]{%
 \@ifx{#1\undefined}
}%
\providecommand \@ifnum [1]{%
 \ifnum #1\expandafter \@firstoftwo
 \else \expandafter \@secondoftwo
 \fi
}%
\providecommand \@ifx [1]{%
 \ifx #1\expandafter \@firstoftwo
 \else \expandafter \@secondoftwo
 \fi
}%
\providecommand \natexlab [1]{#1}%
\providecommand \enquote  [1]{``#1''}%
\providecommand \bibnamefont  [1]{#1}%
\providecommand \bibfnamefont [1]{#1}%
\providecommand \citenamefont [1]{#1}%
\providecommand \href@noop [0]{\@secondoftwo}%
\providecommand \href [0]{\begingroup \@sanitize@url \@href}%
\providecommand \@href[1]{\@@startlink{#1}\@@href}%
\providecommand \@@href[1]{\endgroup#1\@@endlink}%
\providecommand \@sanitize@url [0]{\catcode `\\12\catcode `\$12\catcode
  `\&12\catcode `\#12\catcode `\^12\catcode `\_12\catcode `\%12\relax}%
\providecommand \@@startlink[1]{}%
\providecommand \@@endlink[0]{}%
\providecommand \url  [0]{\begingroup\@sanitize@url \@url }%
\providecommand \@url [1]{\endgroup\@href {#1}{\urlprefix }}%
\providecommand \urlprefix  [0]{URL }%
\providecommand \Eprint [0]{\href }%
\providecommand \doibase [0]{https://doi.org/}%
\providecommand \selectlanguage [0]{\@gobble}%
\providecommand \bibinfo  [0]{\@secondoftwo}%
\providecommand \bibfield  [0]{\@secondoftwo}%
\providecommand \translation [1]{[#1]}%
\providecommand \BibitemOpen [0]{}%
\providecommand \bibitemStop [0]{}%
\providecommand \bibitemNoStop [0]{.\EOS\space}%
\providecommand \EOS [0]{\spacefactor3000\relax}%
\providecommand \BibitemShut  [1]{\csname bibitem#1\endcsname}%
\let\auto@bib@innerbib\@empty
%</preamble>
\bibitem [{\citenamefont {Kokkotas}\ and\ \citenamefont
  {Schmidt}(1999)}]{Kokkotas:1999bd}%
  \BibitemOpen
  \bibfield  {author} {\bibinfo {author} {\bibfnamefont {K.~D.}\ \bibnamefont
  {Kokkotas}}\ and\ \bibinfo {author} {\bibfnamefont {B.~G.}\ \bibnamefont
  {Schmidt}},\ }\bibfield  {title} {\bibinfo {title} {{Quasinormal modes of
  stars and black holes}},\ }\href {https://doi.org/10.12942/lrr-1999-2}
  {\bibfield  {journal} {\bibinfo  {journal} {Living Rev. Rel.}\ }\textbf
  {\bibinfo {volume} {2}},\ \bibinfo {pages} {2} (\bibinfo {year} {1999})},\
  \Eprint {https://arxiv.org/abs/gr-qc/9909058} {arXiv:gr-qc/9909058}
  \BibitemShut {NoStop}%
\bibitem [{\citenamefont {Nollert}(1996)}]{Nollert:1996rf}%
  \BibitemOpen
  \bibfield  {author} {\bibinfo {author} {\bibfnamefont {H.-P.}\ \bibnamefont
  {Nollert}},\ }\bibfield  {title} {\bibinfo {title} {{About the significance
  of quasinormal modes of black holes}},\ }\href
  {https://doi.org/10.1103/PhysRevD.53.4397} {\bibfield  {journal} {\bibinfo
  {journal} {Phys. Rev. D}\ }\textbf {\bibinfo {volume} {53}},\ \bibinfo
  {pages} {4397} (\bibinfo {year} {1996})},\ \Eprint
  {https://arxiv.org/abs/gr-qc/9602032} {arXiv:gr-qc/9602032} \BibitemShut
  {NoStop}%
\bibitem [{\citenamefont {Berti}\ \emph {et~al.}(2009)\citenamefont {Berti},
  \citenamefont {Cardoso},\ and\ \citenamefont {Starinets}}]{Berti:2009kk}%
  \BibitemOpen
  \bibfield  {author} {\bibinfo {author} {\bibfnamefont {E.}~\bibnamefont
  {Berti}}, \bibinfo {author} {\bibfnamefont {V.}~\bibnamefont {Cardoso}},\
  and\ \bibinfo {author} {\bibfnamefont {A.~O.}\ \bibnamefont {Starinets}},\
  }\bibfield  {title} {\bibinfo {title} {{Quasinormal modes of black holes and
  black branes}},\ }\href {https://doi.org/10.1088/0264-9381/26/16/163001}
  {\bibfield  {journal} {\bibinfo  {journal} {Class. Quant. Grav.}\ }\textbf
  {\bibinfo {volume} {26}},\ \bibinfo {pages} {163001} (\bibinfo {year}
  {2009})},\ \Eprint {https://arxiv.org/abs/0905.2975} {arXiv:0905.2975
  [gr-qc]} \BibitemShut {NoStop}%
\bibitem [{\citenamefont {Konoplya}\ and\ \citenamefont
  {Zhidenko}(2011)}]{Konoplya:2011qq}%
  \BibitemOpen
  \bibfield  {author} {\bibinfo {author} {\bibfnamefont {R.~A.}\ \bibnamefont
  {Konoplya}}\ and\ \bibinfo {author} {\bibfnamefont {A.}~\bibnamefont
  {Zhidenko}},\ }\bibfield  {title} {\bibinfo {title} {{Quasinormal modes of
  black holes: From astrophysics to string theory}},\ }\href
  {https://doi.org/10.1103/RevModPhys.83.793} {\bibfield  {journal} {\bibinfo
  {journal} {Rev. Mod. Phys.}\ }\textbf {\bibinfo {volume} {83}},\ \bibinfo
  {pages} {793} (\bibinfo {year} {2011})},\ \Eprint
  {https://arxiv.org/abs/1102.4014} {arXiv:1102.4014 [gr-qc]} \BibitemShut
  {NoStop}%
\bibitem [{\citenamefont {Barausse}\ \emph {et~al.}(2014)\citenamefont
  {Barausse}, \citenamefont {Cardoso},\ and\ \citenamefont
  {Pani}}]{Barausse:2014tra}%
  \BibitemOpen
  \bibfield  {author} {\bibinfo {author} {\bibfnamefont {E.}~\bibnamefont
  {Barausse}}, \bibinfo {author} {\bibfnamefont {V.}~\bibnamefont {Cardoso}},\
  and\ \bibinfo {author} {\bibfnamefont {P.}~\bibnamefont {Pani}},\ }\bibfield
  {title} {\bibinfo {title} {{Can environmental effects spoil precision
  gravitational-wave astrophysics?}},\ }\href
  {https://doi.org/10.1103/PhysRevD.89.104059} {\bibfield  {journal} {\bibinfo
  {journal} {Phys. Rev. D}\ }\textbf {\bibinfo {volume} {89}},\ \bibinfo
  {pages} {104059} (\bibinfo {year} {2014})},\ \Eprint
  {https://arxiv.org/abs/1404.7149} {arXiv:1404.7149 [gr-qc]} \BibitemShut
  {NoStop}%
\bibitem [{\citenamefont {Dreyer}\ \emph {et~al.}(2004)\citenamefont {Dreyer},
  \citenamefont {Kelly}, \citenamefont {Krishnan}, \citenamefont {Finn},
  \citenamefont {Garrison},\ and\ \citenamefont
  {Lopez-Aleman}}]{Dreyer:2003bv}%
  \BibitemOpen
  \bibfield  {author} {\bibinfo {author} {\bibfnamefont {O.}~\bibnamefont
  {Dreyer}}, \bibinfo {author} {\bibfnamefont {B.~J.}\ \bibnamefont {Kelly}},
  \bibinfo {author} {\bibfnamefont {B.}~\bibnamefont {Krishnan}}, \bibinfo
  {author} {\bibfnamefont {L.~S.}\ \bibnamefont {Finn}}, \bibinfo {author}
  {\bibfnamefont {D.}~\bibnamefont {Garrison}},\ and\ \bibinfo {author}
  {\bibfnamefont {R.}~\bibnamefont {Lopez-Aleman}},\ }\bibfield  {title}
  {\bibinfo {title} {{Black hole spectroscopy: Testing general relativity
  through gravitational wave observations}},\ }\href
  {https://doi.org/10.1088/0264-9381/21/4/003} {\bibfield  {journal} {\bibinfo
  {journal} {Class. Quant. Grav.}\ }\textbf {\bibinfo {volume} {21}},\ \bibinfo
  {pages} {787} (\bibinfo {year} {2004})},\ \Eprint
  {https://arxiv.org/abs/gr-qc/0309007} {arXiv:gr-qc/0309007 [gr-qc]}
  \BibitemShut {NoStop}%
%%CITATION = GR-QC/0309007;%%
\bibitem [{\citenamefont {Berti}\ \emph {et~al.}(2006)\citenamefont {Berti},
  \citenamefont {Cardoso},\ and\ \citenamefont {Will}}]{Berti:2005ys}%
  \BibitemOpen
  \bibfield  {author} {\bibinfo {author} {\bibfnamefont {E.}~\bibnamefont
  {Berti}}, \bibinfo {author} {\bibfnamefont {V.}~\bibnamefont {Cardoso}},\
  and\ \bibinfo {author} {\bibfnamefont {C.~M.}\ \bibnamefont {Will}},\
  }\bibfield  {title} {\bibinfo {title} {{On gravitational-wave spectroscopy of
  massive black holes with the space interferometer LISA}},\ }\href
  {https://doi.org/10.1103/PhysRevD.73.064030} {\bibfield  {journal} {\bibinfo
  {journal} {Phys. Rev. D}\ }\textbf {\bibinfo {volume} {73}},\ \bibinfo
  {pages} {064030} (\bibinfo {year} {2006})},\ \Eprint
  {https://arxiv.org/abs/gr-qc/0512160} {arXiv:gr-qc/0512160} \BibitemShut
  {NoStop}%
\bibitem [{\citenamefont {Berti}\ \emph {et~al.}(2016)\citenamefont {Berti},
  \citenamefont {Sesana}, \citenamefont {Barausse}, \citenamefont {Cardoso},\
  and\ \citenamefont {Belczynski}}]{Berti:2016lat}%
  \BibitemOpen
  \bibfield  {author} {\bibinfo {author} {\bibfnamefont {E.}~\bibnamefont
  {Berti}}, \bibinfo {author} {\bibfnamefont {A.}~\bibnamefont {Sesana}},
  \bibinfo {author} {\bibfnamefont {E.}~\bibnamefont {Barausse}}, \bibinfo
  {author} {\bibfnamefont {V.}~\bibnamefont {Cardoso}},\ and\ \bibinfo {author}
  {\bibfnamefont {K.}~\bibnamefont {Belczynski}},\ }\bibfield  {title}
  {\bibinfo {title} {{Spectroscopy of Kerr black holes with Earth- and
  space-based interferometers}},\ }\href
  {https://doi.org/10.1103/PhysRevLett.117.101102} {\bibfield  {journal}
  {\bibinfo  {journal} {Phys. Rev. Lett.}\ }\textbf {\bibinfo {volume} {117}},\
  \bibinfo {pages} {101102} (\bibinfo {year} {2016})},\ \Eprint
  {https://arxiv.org/abs/1605.09286} {arXiv:1605.09286 [gr-qc]} \BibitemShut
  {NoStop}%
\bibitem [{\citenamefont {Abbott}\ \emph {et~al.}(2016)\citenamefont {Abbott}
  \emph {et~al.}}]{LIGOScientific:2016aoc}%
  \BibitemOpen
  \bibfield  {author} {\bibinfo {author} {\bibfnamefont {B.~P.}\ \bibnamefont
  {Abbott}} \emph {et~al.} (\bibinfo {collaboration} {LIGO Scientific,
  Virgo}),\ }\bibfield  {title} {\bibinfo {title} {{Observation of
  Gravitational Waves from a Binary Black Hole Merger}},\ }\href
  {https://doi.org/10.1103/PhysRevLett.116.061102} {\bibfield  {journal}
  {\bibinfo  {journal} {Phys. Rev. Lett.}\ }\textbf {\bibinfo {volume} {116}},\
  \bibinfo {pages} {061102} (\bibinfo {year} {2016})},\ \Eprint
  {https://arxiv.org/abs/1602.03837} {arXiv:1602.03837 [gr-qc]} \BibitemShut
  {NoStop}%
\bibitem [{\citenamefont {Abbott}\ \emph
  {et~al.}(2021{\natexlab{a}})\citenamefont {Abbott} \emph
  {et~al.}}]{LIGOScientific:2020tif}%
  \BibitemOpen
  \bibfield  {author} {\bibinfo {author} {\bibfnamefont {R.}~\bibnamefont
  {Abbott}} \emph {et~al.} (\bibinfo {collaboration} {LIGO Scientific,
  Virgo}),\ }\bibfield  {title} {\bibinfo {title} {{Tests of general relativity
  with binary black holes from the second LIGO-Virgo gravitational-wave
  transient catalog}},\ }\href {https://doi.org/10.1103/PhysRevD.103.122002}
  {\bibfield  {journal} {\bibinfo  {journal} {Phys. Rev. D}\ }\textbf {\bibinfo
  {volume} {103}},\ \bibinfo {pages} {122002} (\bibinfo {year}
  {2021}{\natexlab{a}})},\ \Eprint {https://arxiv.org/abs/2010.14529}
  {arXiv:2010.14529 [gr-qc]} \BibitemShut {NoStop}%
\bibitem [{\citenamefont {Abbott}\ \emph
  {et~al.}(2021{\natexlab{b}})\citenamefont {Abbott} \emph
  {et~al.}}]{LIGOScientific:2021sio}%
  \BibitemOpen
  \bibfield  {author} {\bibinfo {author} {\bibfnamefont {R.}~\bibnamefont
  {Abbott}} \emph {et~al.} (\bibinfo {collaboration} {LIGO Scientific, VIRGO,
  KAGRA}),\ }\bibfield  {title} {\bibinfo {title} {{Tests of General Relativity
  with GWTC-3}},\ }\href@noop {} {\  (\bibinfo {year} {2021}{\natexlab{b}})},\
  \Eprint {https://arxiv.org/abs/2112.06861} {arXiv:2112.06861 [gr-qc]}
  \BibitemShut {NoStop}%
\bibitem [{\citenamefont {Isi}\ \emph {et~al.}(2019)\citenamefont {Isi},
  \citenamefont {Giesler}, \citenamefont {Farr}, \citenamefont {Scheel},\ and\
  \citenamefont {Teukolsky}}]{Isi:2019aib}%
  \BibitemOpen
  \bibfield  {author} {\bibinfo {author} {\bibfnamefont {M.}~\bibnamefont
  {Isi}}, \bibinfo {author} {\bibfnamefont {M.}~\bibnamefont {Giesler}},
  \bibinfo {author} {\bibfnamefont {W.~M.}\ \bibnamefont {Farr}}, \bibinfo
  {author} {\bibfnamefont {M.~A.}\ \bibnamefont {Scheel}},\ and\ \bibinfo
  {author} {\bibfnamefont {S.~A.}\ \bibnamefont {Teukolsky}},\ }\bibfield
  {title} {\bibinfo {title} {{Testing the no-hair theorem with GW150914}},\
  }\href {https://doi.org/10.1103/PhysRevLett.123.111102} {\bibfield  {journal}
  {\bibinfo  {journal} {Phys. Rev. Lett.}\ }\textbf {\bibinfo {volume} {123}},\
  \bibinfo {pages} {111102} (\bibinfo {year} {2019})},\ \Eprint
  {https://arxiv.org/abs/1905.00869} {arXiv:1905.00869 [gr-qc]} \BibitemShut
  {NoStop}%
\bibitem [{\citenamefont {Capano}\ and\ \citenamefont
  {Nitz}(2020)}]{Capano:2020dix}%
  \BibitemOpen
  \bibfield  {author} {\bibinfo {author} {\bibfnamefont {C.~D.}\ \bibnamefont
  {Capano}}\ and\ \bibinfo {author} {\bibfnamefont {A.~H.}\ \bibnamefont
  {Nitz}},\ }\bibfield  {title} {\bibinfo {title} {{Binary black hole
  spectroscopy: a no-hair test of GW190814 and GW190412}},\ }\href
  {https://doi.org/10.1103/PhysRevD.102.124070} {\bibfield  {journal} {\bibinfo
   {journal} {Phys. Rev. D}\ }\textbf {\bibinfo {volume} {102}},\ \bibinfo
  {pages} {124070} (\bibinfo {year} {2020})},\ \Eprint
  {https://arxiv.org/abs/2008.02248} {arXiv:2008.02248 [gr-qc]} \BibitemShut
  {NoStop}%
\bibitem [{\citenamefont {Capano}\ \emph {et~al.}(2023)\citenamefont {Capano},
  \citenamefont {Cabero}, \citenamefont {Westerweck}, \citenamefont {Abedi},
  \citenamefont {Kastha}, \citenamefont {Nitz}, \citenamefont {Wang},
  \citenamefont {Nielsen},\ and\ \citenamefont {Krishnan}}]{Capano:2021etf}%
  \BibitemOpen
  \bibfield  {author} {\bibinfo {author} {\bibfnamefont {C.~D.}\ \bibnamefont
  {Capano}}, \bibinfo {author} {\bibfnamefont {M.}~\bibnamefont {Cabero}},
  \bibinfo {author} {\bibfnamefont {J.}~\bibnamefont {Westerweck}}, \bibinfo
  {author} {\bibfnamefont {J.}~\bibnamefont {Abedi}}, \bibinfo {author}
  {\bibfnamefont {S.}~\bibnamefont {Kastha}}, \bibinfo {author} {\bibfnamefont
  {A.~H.}\ \bibnamefont {Nitz}}, \bibinfo {author} {\bibfnamefont {Y.-F.}\
  \bibnamefont {Wang}}, \bibinfo {author} {\bibfnamefont {A.~B.}\ \bibnamefont
  {Nielsen}},\ and\ \bibinfo {author} {\bibfnamefont {B.}~\bibnamefont
  {Krishnan}},\ }\bibfield  {title} {\bibinfo {title} {{Multimode Quasinormal
  Spectrum from a Perturbed Black Hole}},\ }\href
  {https://doi.org/10.1103/PhysRevLett.131.221402} {\bibfield  {journal}
  {\bibinfo  {journal} {Phys. Rev. Lett.}\ }\textbf {\bibinfo {volume} {131}},\
  \bibinfo {pages} {221402} (\bibinfo {year} {2023})},\ \Eprint
  {https://arxiv.org/abs/2105.05238} {arXiv:2105.05238 [gr-qc]} \BibitemShut
  {NoStop}%
\bibitem [{\citenamefont {Cotesta}\ \emph {et~al.}(2022)\citenamefont
  {Cotesta}, \citenamefont {Carullo}, \citenamefont {Berti},\ and\
  \citenamefont {Cardoso}}]{Cotesta:2022pci}%
  \BibitemOpen
  \bibfield  {author} {\bibinfo {author} {\bibfnamefont {R.}~\bibnamefont
  {Cotesta}}, \bibinfo {author} {\bibfnamefont {G.}~\bibnamefont {Carullo}},
  \bibinfo {author} {\bibfnamefont {E.}~\bibnamefont {Berti}},\ and\ \bibinfo
  {author} {\bibfnamefont {V.}~\bibnamefont {Cardoso}},\ }\bibfield  {title}
  {\bibinfo {title} {{Analysis of Ringdown Overtones in GW150914}},\ }\href
  {https://doi.org/10.1103/PhysRevLett.129.111102} {\bibfield  {journal}
  {\bibinfo  {journal} {Phys. Rev. Lett.}\ }\textbf {\bibinfo {volume} {129}},\
  \bibinfo {pages} {111102} (\bibinfo {year} {2022})},\ \Eprint
  {https://arxiv.org/abs/2201.00822} {arXiv:2201.00822 [gr-qc]} \BibitemShut
  {NoStop}%
\bibitem [{\citenamefont {Capano}\ \emph {et~al.}(2024)\citenamefont {Capano},
  \citenamefont {Abedi}, \citenamefont {Kastha}, \citenamefont {Nitz},
  \citenamefont {Westerweck}, \citenamefont {Wang}, \citenamefont {Cabero},
  \citenamefont {Nielsen},\ and\ \citenamefont {Krishnan}}]{Capano:2022zqm}%
  \BibitemOpen
  \bibfield  {author} {\bibinfo {author} {\bibfnamefont {C.~D.}\ \bibnamefont
  {Capano}}, \bibinfo {author} {\bibfnamefont {J.}~\bibnamefont {Abedi}},
  \bibinfo {author} {\bibfnamefont {S.}~\bibnamefont {Kastha}}, \bibinfo
  {author} {\bibfnamefont {A.~H.}\ \bibnamefont {Nitz}}, \bibinfo {author}
  {\bibfnamefont {J.}~\bibnamefont {Westerweck}}, \bibinfo {author}
  {\bibfnamefont {Y.-F.}\ \bibnamefont {Wang}}, \bibinfo {author}
  {\bibfnamefont {M.}~\bibnamefont {Cabero}}, \bibinfo {author} {\bibfnamefont
  {A.~B.}\ \bibnamefont {Nielsen}},\ and\ \bibinfo {author} {\bibfnamefont
  {B.}~\bibnamefont {Krishnan}},\ }\bibfield  {title} {\bibinfo {title}
  {{Estimating false alarm rates of sub-dominant quasi-normal modes in
  GW190521}},\ }\href {https://doi.org/10.1088/1361-6382/ad84ae} {\bibfield
  {journal} {\bibinfo  {journal} {Class. Quant. Grav.}\ }\textbf {\bibinfo
  {volume} {41}},\ \bibinfo {pages} {245009} (\bibinfo {year} {2024})},\
  \Eprint {https://arxiv.org/abs/2209.00640} {arXiv:2209.00640 [gr-qc]}
  \BibitemShut {NoStop}%
\bibitem [{\citenamefont {Forteza}\ \emph {et~al.}(2023)\citenamefont
  {Forteza}, \citenamefont {Bhagwat}, \citenamefont {Kumar},\ and\
  \citenamefont {Pani}}]{Forteza:2022tgq}%
  \BibitemOpen
  \bibfield  {author} {\bibinfo {author} {\bibfnamefont {X.~J.}\ \bibnamefont
  {Forteza}}, \bibinfo {author} {\bibfnamefont {S.}~\bibnamefont {Bhagwat}},
  \bibinfo {author} {\bibfnamefont {S.}~\bibnamefont {Kumar}},\ and\ \bibinfo
  {author} {\bibfnamefont {P.}~\bibnamefont {Pani}},\ }\bibfield  {title}
  {\bibinfo {title} {{Novel Ringdown Amplitude-Phase Consistency Test}},\
  }\href {https://doi.org/10.1103/PhysRevLett.130.021001} {\bibfield  {journal}
  {\bibinfo  {journal} {Phys. Rev. Lett.}\ }\textbf {\bibinfo {volume} {130}},\
  \bibinfo {pages} {021001} (\bibinfo {year} {2023})},\ \Eprint
  {https://arxiv.org/abs/2205.14910} {arXiv:2205.14910 [gr-qc]} \BibitemShut
  {NoStop}%
\bibitem [{\citenamefont {Finch}\ and\ \citenamefont
  {Moore}(2022)}]{Finch:2022ynt}%
  \BibitemOpen
  \bibfield  {author} {\bibinfo {author} {\bibfnamefont {E.}~\bibnamefont
  {Finch}}\ and\ \bibinfo {author} {\bibfnamefont {C.~J.}\ \bibnamefont
  {Moore}},\ }\bibfield  {title} {\bibinfo {title} {{Searching for a ringdown
  overtone in GW150914}},\ }\href {https://doi.org/10.1103/PhysRevD.106.043005}
  {\bibfield  {journal} {\bibinfo  {journal} {Phys. Rev. D}\ }\textbf {\bibinfo
  {volume} {106}},\ \bibinfo {pages} {043005} (\bibinfo {year} {2022})},\
  \Eprint {https://arxiv.org/abs/2205.07809} {arXiv:2205.07809 [gr-qc]}
  \BibitemShut {NoStop}%
\bibitem [{\citenamefont {Abedi}\ \emph {et~al.}(2023)\citenamefont {Abedi},
  \citenamefont {Capano}, \citenamefont {Kastha}, \citenamefont {Nitz},
  \citenamefont {Wang}, \citenamefont {Westerweck}, \citenamefont {Nielsen},\
  and\ \citenamefont {Krishnan}}]{Abedi:2023kot}%
  \BibitemOpen
  \bibfield  {author} {\bibinfo {author} {\bibfnamefont {J.}~\bibnamefont
  {Abedi}}, \bibinfo {author} {\bibfnamefont {C.~D.}\ \bibnamefont {Capano}},
  \bibinfo {author} {\bibfnamefont {S.}~\bibnamefont {Kastha}}, \bibinfo
  {author} {\bibfnamefont {A.~H.}\ \bibnamefont {Nitz}}, \bibinfo {author}
  {\bibfnamefont {Y.-F.}\ \bibnamefont {Wang}}, \bibinfo {author}
  {\bibfnamefont {J.}~\bibnamefont {Westerweck}}, \bibinfo {author}
  {\bibfnamefont {A.~B.}\ \bibnamefont {Nielsen}},\ and\ \bibinfo {author}
  {\bibfnamefont {B.}~\bibnamefont {Krishnan}},\ }\bibfield  {title} {\bibinfo
  {title} {{Spectroscopy for asymmetric binary black hole mergers}},\ }\href
  {https://doi.org/10.1103/PhysRevD.108.104009} {\bibfield  {journal} {\bibinfo
   {journal} {Phys. Rev. D}\ }\textbf {\bibinfo {volume} {108}},\ \bibinfo
  {pages} {104009} (\bibinfo {year} {2023})},\ \Eprint
  {https://arxiv.org/abs/2309.03121} {arXiv:2309.03121 [gr-qc]} \BibitemShut
  {NoStop}%
\bibitem [{\citenamefont {Carullo}\ \emph {et~al.}(2023)\citenamefont
  {Carullo}, \citenamefont {Cotesta}, \citenamefont {Berti},\ and\
  \citenamefont {Cardoso}}]{Carullo:2023gtf}%
  \BibitemOpen
  \bibfield  {author} {\bibinfo {author} {\bibfnamefont {G.}~\bibnamefont
  {Carullo}}, \bibinfo {author} {\bibfnamefont {R.}~\bibnamefont {Cotesta}},
  \bibinfo {author} {\bibfnamefont {E.}~\bibnamefont {Berti}},\ and\ \bibinfo
  {author} {\bibfnamefont {V.}~\bibnamefont {Cardoso}},\ }\bibfield  {title}
  {\bibinfo {title} {{Reply to Comment on ''Analysis of Ringdown Overtones in
  GW150914''}},\ }\href {https://doi.org/10.1103/PhysRevLett.131.169002}
  {\bibfield  {journal} {\bibinfo  {journal} {Phys. Rev. Lett.}\ }\textbf
  {\bibinfo {volume} {131}},\ \bibinfo {pages} {169002} (\bibinfo {year}
  {2023})},\ \Eprint {https://arxiv.org/abs/2310.20625} {arXiv:2310.20625
  [gr-qc]} \BibitemShut {NoStop}%
\bibitem [{\citenamefont {Baibhav}\ \emph {et~al.}(2023)\citenamefont
  {Baibhav}, \citenamefont {Cheung}, \citenamefont {Berti}, \citenamefont
  {Cardoso}, \citenamefont {Carullo}, \citenamefont {Cotesta}, \citenamefont
  {Del~Pozzo},\ and\ \citenamefont {Duque}}]{Baibhav:2023clw}%
  \BibitemOpen
  \bibfield  {author} {\bibinfo {author} {\bibfnamefont {V.}~\bibnamefont
  {Baibhav}}, \bibinfo {author} {\bibfnamefont {M.~H.-Y.}\ \bibnamefont
  {Cheung}}, \bibinfo {author} {\bibfnamefont {E.}~\bibnamefont {Berti}},
  \bibinfo {author} {\bibfnamefont {V.}~\bibnamefont {Cardoso}}, \bibinfo
  {author} {\bibfnamefont {G.}~\bibnamefont {Carullo}}, \bibinfo {author}
  {\bibfnamefont {R.}~\bibnamefont {Cotesta}}, \bibinfo {author} {\bibfnamefont
  {W.}~\bibnamefont {Del~Pozzo}},\ and\ \bibinfo {author} {\bibfnamefont
  {F.}~\bibnamefont {Duque}},\ }\bibfield  {title} {\bibinfo {title} {{Agnostic
  black hole spectroscopy: Quasinormal mode content of numerical relativity
  waveforms and limits of validity of linear perturbation theory}},\ }\href
  {https://doi.org/10.1103/PhysRevD.108.104020} {\bibfield  {journal} {\bibinfo
   {journal} {Phys. Rev. D}\ }\textbf {\bibinfo {volume} {108}},\ \bibinfo
  {pages} {104020} (\bibinfo {year} {2023})},\ \Eprint
  {https://arxiv.org/abs/2302.03050} {arXiv:2302.03050 [gr-qc]} \BibitemShut
  {NoStop}%
\bibitem [{\citenamefont {Nee}\ \emph {et~al.}(2023)\citenamefont {Nee},
  \citenamefont {V\"olkel},\ and\ \citenamefont {Pfeiffer}}]{Nee:2023osy}%
  \BibitemOpen
  \bibfield  {author} {\bibinfo {author} {\bibfnamefont {P.~J.}\ \bibnamefont
  {Nee}}, \bibinfo {author} {\bibfnamefont {S.~H.}\ \bibnamefont {V\"olkel}},\
  and\ \bibinfo {author} {\bibfnamefont {H.~P.}\ \bibnamefont {Pfeiffer}},\
  }\bibfield  {title} {\bibinfo {title} {{Role of black hole quasinormal mode
  overtones for ringdown analysis}},\ }\href
  {https://doi.org/10.1103/PhysRevD.108.044032} {\bibfield  {journal} {\bibinfo
   {journal} {Phys. Rev. D}\ }\textbf {\bibinfo {volume} {108}},\ \bibinfo
  {pages} {044032} (\bibinfo {year} {2023})},\ \Eprint
  {https://arxiv.org/abs/2302.06634} {arXiv:2302.06634 [gr-qc]} \BibitemShut
  {NoStop}%
\bibitem [{\citenamefont {Zhu}\ \emph {et~al.}(2024{\natexlab{a}})\citenamefont
  {Zhu}, \citenamefont {Ripley}, \citenamefont {C\'ardenas-Avenda\~no},\ and\
  \citenamefont {Pretorius}}]{Zhu:2023mzv}%
  \BibitemOpen
  \bibfield  {author} {\bibinfo {author} {\bibfnamefont {H.}~\bibnamefont
  {Zhu}}, \bibinfo {author} {\bibfnamefont {J.~L.}\ \bibnamefont {Ripley}},
  \bibinfo {author} {\bibfnamefont {A.}~\bibnamefont {C\'ardenas-Avenda\~no}},\
  and\ \bibinfo {author} {\bibfnamefont {F.}~\bibnamefont {Pretorius}},\
  }\bibfield  {title} {\bibinfo {title} {{Challenges in quasinormal mode
  extraction: Perspectives from numerical solutions to the Teukolsky
  equation}},\ }\href {https://doi.org/10.1103/PhysRevD.109.044010} {\bibfield
  {journal} {\bibinfo  {journal} {Phys. Rev. D}\ }\textbf {\bibinfo {volume}
  {109}},\ \bibinfo {pages} {044010} (\bibinfo {year} {2024}{\natexlab{a}})},\
  \Eprint {https://arxiv.org/abs/2309.13204} {arXiv:2309.13204 [gr-qc]}
  \BibitemShut {NoStop}%
\bibitem [{\citenamefont {Siegel}\ \emph {et~al.}(2023)\citenamefont {Siegel},
  \citenamefont {Isi},\ and\ \citenamefont {Farr}}]{Siegel:2023lxl}%
  \BibitemOpen
  \bibfield  {author} {\bibinfo {author} {\bibfnamefont {H.}~\bibnamefont
  {Siegel}}, \bibinfo {author} {\bibfnamefont {M.}~\bibnamefont {Isi}},\ and\
  \bibinfo {author} {\bibfnamefont {W.~M.}\ \bibnamefont {Farr}},\ }\bibfield
  {title} {\bibinfo {title} {{Ringdown of GW190521: Hints of multiple
  quasinormal modes with a precessional interpretation}},\ }\href
  {https://doi.org/10.1103/PhysRevD.108.064008} {\bibfield  {journal} {\bibinfo
   {journal} {Phys. Rev. D}\ }\textbf {\bibinfo {volume} {108}},\ \bibinfo
  {pages} {064008} (\bibinfo {year} {2023})},\ \Eprint
  {https://arxiv.org/abs/2307.11975} {arXiv:2307.11975 [gr-qc]} \BibitemShut
  {NoStop}%
\bibitem [{\citenamefont {Gennari}\ \emph {et~al.}(2024)\citenamefont
  {Gennari}, \citenamefont {Carullo},\ and\ \citenamefont
  {Del~Pozzo}}]{Gennari:2023gmx}%
  \BibitemOpen
  \bibfield  {author} {\bibinfo {author} {\bibfnamefont {V.}~\bibnamefont
  {Gennari}}, \bibinfo {author} {\bibfnamefont {G.}~\bibnamefont {Carullo}},\
  and\ \bibinfo {author} {\bibfnamefont {W.}~\bibnamefont {Del~Pozzo}},\
  }\bibfield  {title} {\bibinfo {title} {{Searching for ringdown higher modes
  with a numerical relativity-informed post-merger model}},\ }\href
  {https://doi.org/10.1140/epjc/s10052-024-12550-x} {\bibfield  {journal}
  {\bibinfo  {journal} {Eur. Phys. J. C}\ }\textbf {\bibinfo {volume} {84}},\
  \bibinfo {pages} {233} (\bibinfo {year} {2024})},\ \Eprint
  {https://arxiv.org/abs/2312.12515} {arXiv:2312.12515 [gr-qc]} \BibitemShut
  {NoStop}%
\bibitem [{\citenamefont {Maggiore}\ \emph {et~al.}(2020)\citenamefont
  {Maggiore} \emph {et~al.}}]{Maggiore:2019uih}%
  \BibitemOpen
  \bibfield  {author} {\bibinfo {author} {\bibfnamefont {M.}~\bibnamefont
  {Maggiore}} \emph {et~al.},\ }\bibfield  {title} {\bibinfo {title} {{Science
  Case for the Einstein Telescope}},\ }\href
  {https://doi.org/10.1088/1475-7516/2020/03/050} {\bibfield  {journal}
  {\bibinfo  {journal} {JCAP}\ }\textbf {\bibinfo {volume} {03}},\ \bibinfo
  {pages} {050}},\ \Eprint {https://arxiv.org/abs/1912.02622} {arXiv:1912.02622
  [astro-ph.CO]} \BibitemShut {NoStop}%
\bibitem [{\citenamefont {Cabero}\ \emph {et~al.}(2020)\citenamefont {Cabero},
  \citenamefont {Westerweck}, \citenamefont {Capano}, \citenamefont {Kumar},
  \citenamefont {Nielsen},\ and\ \citenamefont {Krishnan}}]{Cabero:2019zyt}%
  \BibitemOpen
  \bibfield  {author} {\bibinfo {author} {\bibfnamefont {M.}~\bibnamefont
  {Cabero}}, \bibinfo {author} {\bibfnamefont {J.}~\bibnamefont {Westerweck}},
  \bibinfo {author} {\bibfnamefont {C.~D.}\ \bibnamefont {Capano}}, \bibinfo
  {author} {\bibfnamefont {S.}~\bibnamefont {Kumar}}, \bibinfo {author}
  {\bibfnamefont {A.~B.}\ \bibnamefont {Nielsen}},\ and\ \bibinfo {author}
  {\bibfnamefont {B.}~\bibnamefont {Krishnan}},\ }\bibfield  {title} {\bibinfo
  {title} {{Black hole spectroscopy in the next decade}},\ }\href
  {https://doi.org/10.1103/PhysRevD.101.064044} {\bibfield  {journal} {\bibinfo
   {journal} {Phys. Rev. D}\ }\textbf {\bibinfo {volume} {101}},\ \bibinfo
  {pages} {064044} (\bibinfo {year} {2020})},\ \Eprint
  {https://arxiv.org/abs/1911.01361} {arXiv:1911.01361 [gr-qc]} \BibitemShut
  {NoStop}%
\bibitem [{\citenamefont {Toubiana}\ \emph {et~al.}(2024)\citenamefont
  {Toubiana}, \citenamefont {Pompili}, \citenamefont {Buonanno}, \citenamefont
  {Gair},\ and\ \citenamefont {Katz}}]{Toubiana:2023cwr}%
  \BibitemOpen
  \bibfield  {author} {\bibinfo {author} {\bibfnamefont {A.}~\bibnamefont
  {Toubiana}}, \bibinfo {author} {\bibfnamefont {L.}~\bibnamefont {Pompili}},
  \bibinfo {author} {\bibfnamefont {A.}~\bibnamefont {Buonanno}}, \bibinfo
  {author} {\bibfnamefont {J.~R.}\ \bibnamefont {Gair}},\ and\ \bibinfo
  {author} {\bibfnamefont {M.~L.}\ \bibnamefont {Katz}},\ }\bibfield  {title}
  {\bibinfo {title} {{Measuring source properties and quasinormal mode
  frequencies of heavy massive black-hole binaries with LISA}},\ }\href
  {https://doi.org/10.1103/PhysRevD.109.104019} {\bibfield  {journal} {\bibinfo
   {journal} {Phys. Rev. D}\ }\textbf {\bibinfo {volume} {109}},\ \bibinfo
  {pages} {104019} (\bibinfo {year} {2024})},\ \Eprint
  {https://arxiv.org/abs/2307.15086} {arXiv:2307.15086 [gr-qc]} \BibitemShut
  {NoStop}%
\bibitem [{\citenamefont {London}\ \emph {et~al.}(2014)\citenamefont {London},
  \citenamefont {Shoemaker},\ and\ \citenamefont {Healy}}]{London:2014cma}%
  \BibitemOpen
  \bibfield  {author} {\bibinfo {author} {\bibfnamefont {L.}~\bibnamefont
  {London}}, \bibinfo {author} {\bibfnamefont {D.}~\bibnamefont {Shoemaker}},\
  and\ \bibinfo {author} {\bibfnamefont {J.}~\bibnamefont {Healy}},\ }\bibfield
   {title} {\bibinfo {title} {{Modeling ringdown: Beyond the fundamental
  quasinormal modes}},\ }\href {https://doi.org/10.1103/PhysRevD.90.124032}
  {\bibfield  {journal} {\bibinfo  {journal} {Phys. Rev. D}\ }\textbf {\bibinfo
  {volume} {90}},\ \bibinfo {pages} {124032} (\bibinfo {year} {2014})},\
  \bibinfo {note} {[Erratum: Phys.Rev.D 94, 069902 (2016)]},\ \Eprint
  {https://arxiv.org/abs/1404.3197} {arXiv:1404.3197 [gr-qc]} \BibitemShut
  {NoStop}%
\bibitem [{\citenamefont {Cheung}\ \emph {et~al.}(2023)\citenamefont {Cheung}
  \emph {et~al.}}]{Cheung:2022rbm}%
  \BibitemOpen
  \bibfield  {author} {\bibinfo {author} {\bibfnamefont {M.~H.-Y.}\
  \bibnamefont {Cheung}} \emph {et~al.},\ }\bibfield  {title} {\bibinfo {title}
  {{Nonlinear Effects in Black Hole Ringdown}},\ }\href
  {https://doi.org/10.1103/PhysRevLett.130.081401} {\bibfield  {journal}
  {\bibinfo  {journal} {Phys. Rev. Lett.}\ }\textbf {\bibinfo {volume} {130}},\
  \bibinfo {pages} {081401} (\bibinfo {year} {2023})},\ \Eprint
  {https://arxiv.org/abs/2208.07374} {arXiv:2208.07374 [gr-qc]} \BibitemShut
  {NoStop}%
\bibitem [{\citenamefont {Mitman}\ \emph {et~al.}(2023)\citenamefont {Mitman}
  \emph {et~al.}}]{Mitman:2022qdl}%
  \BibitemOpen
  \bibfield  {author} {\bibinfo {author} {\bibfnamefont {K.}~\bibnamefont
  {Mitman}} \emph {et~al.},\ }\bibfield  {title} {\bibinfo {title}
  {{Nonlinearities in Black Hole Ringdowns}},\ }\href
  {https://doi.org/10.1103/PhysRevLett.130.081402} {\bibfield  {journal}
  {\bibinfo  {journal} {Phys. Rev. Lett.}\ }\textbf {\bibinfo {volume} {130}},\
  \bibinfo {pages} {081402} (\bibinfo {year} {2023})},\ \Eprint
  {https://arxiv.org/abs/2208.07380} {arXiv:2208.07380 [gr-qc]} \BibitemShut
  {NoStop}%
\bibitem [{\citenamefont {Zlochower}\ \emph {et~al.}(2003)\citenamefont
  {Zlochower}, \citenamefont {Gomez}, \citenamefont {Husa}, \citenamefont
  {Lehner},\ and\ \citenamefont {Winicour}}]{Zlochower:2003yh}%
  \BibitemOpen
  \bibfield  {author} {\bibinfo {author} {\bibfnamefont {Y.}~\bibnamefont
  {Zlochower}}, \bibinfo {author} {\bibfnamefont {R.}~\bibnamefont {Gomez}},
  \bibinfo {author} {\bibfnamefont {S.}~\bibnamefont {Husa}}, \bibinfo {author}
  {\bibfnamefont {L.}~\bibnamefont {Lehner}},\ and\ \bibinfo {author}
  {\bibfnamefont {J.}~\bibnamefont {Winicour}},\ }\bibfield  {title} {\bibinfo
  {title} {{Mode coupling in the nonlinear response of black holes}},\ }\href
  {https://doi.org/10.1103/PhysRevD.68.084014} {\bibfield  {journal} {\bibinfo
  {journal} {Phys. Rev. D}\ }\textbf {\bibinfo {volume} {68}},\ \bibinfo
  {pages} {084014} (\bibinfo {year} {2003})},\ \Eprint
  {https://arxiv.org/abs/gr-qc/0306098} {arXiv:gr-qc/0306098} \BibitemShut
  {NoStop}%
\bibitem [{\citenamefont {Sberna}\ \emph {et~al.}(2022)\citenamefont {Sberna},
  \citenamefont {Bosch}, \citenamefont {East}, \citenamefont {Green},\ and\
  \citenamefont {Lehner}}]{Sberna:2021eui}%
  \BibitemOpen
  \bibfield  {author} {\bibinfo {author} {\bibfnamefont {L.}~\bibnamefont
  {Sberna}}, \bibinfo {author} {\bibfnamefont {P.}~\bibnamefont {Bosch}},
  \bibinfo {author} {\bibfnamefont {W.~E.}\ \bibnamefont {East}}, \bibinfo
  {author} {\bibfnamefont {S.~R.}\ \bibnamefont {Green}},\ and\ \bibinfo
  {author} {\bibfnamefont {L.}~\bibnamefont {Lehner}},\ }\bibfield  {title}
  {\bibinfo {title} {{Nonlinear effects in the black hole ringdown:
  Absorption-induced mode excitation}},\ }\href
  {https://doi.org/10.1103/PhysRevD.105.064046} {\bibfield  {journal} {\bibinfo
   {journal} {Phys. Rev. D}\ }\textbf {\bibinfo {volume} {105}},\ \bibinfo
  {pages} {064046} (\bibinfo {year} {2022})},\ \Eprint
  {https://arxiv.org/abs/2112.11168} {arXiv:2112.11168 [gr-qc]} \BibitemShut
  {NoStop}%
\bibitem [{\citenamefont {Redondo-Yuste}\ \emph {et~al.}(2024)\citenamefont
  {Redondo-Yuste}, \citenamefont {Carullo}, \citenamefont {Ripley},
  \citenamefont {Berti},\ and\ \citenamefont
  {Cardoso}}]{Redondo-Yuste:2023seq}%
  \BibitemOpen
  \bibfield  {author} {\bibinfo {author} {\bibfnamefont {J.}~\bibnamefont
  {Redondo-Yuste}}, \bibinfo {author} {\bibfnamefont {G.}~\bibnamefont
  {Carullo}}, \bibinfo {author} {\bibfnamefont {J.~L.}\ \bibnamefont {Ripley}},
  \bibinfo {author} {\bibfnamefont {E.}~\bibnamefont {Berti}},\ and\ \bibinfo
  {author} {\bibfnamefont {V.}~\bibnamefont {Cardoso}},\ }\bibfield  {title}
  {\bibinfo {title} {{Spin dependence of black hole ringdown nonlinearities}},\
  }\href {https://doi.org/10.1103/PhysRevD.109.L101503} {\bibfield  {journal}
  {\bibinfo  {journal} {Phys. Rev. D}\ }\textbf {\bibinfo {volume} {109}},\
  \bibinfo {pages} {L101503} (\bibinfo {year} {2024})},\ \Eprint
  {https://arxiv.org/abs/2308.14796} {arXiv:2308.14796 [gr-qc]} \BibitemShut
  {NoStop}%
\bibitem [{\citenamefont {Cheung}\ \emph {et~al.}(2024)\citenamefont {Cheung},
  \citenamefont {Berti}, \citenamefont {Baibhav},\ and\ \citenamefont
  {Cotesta}}]{Cheung:2023vki}%
  \BibitemOpen
  \bibfield  {author} {\bibinfo {author} {\bibfnamefont {M.~H.-Y.}\
  \bibnamefont {Cheung}}, \bibinfo {author} {\bibfnamefont {E.}~\bibnamefont
  {Berti}}, \bibinfo {author} {\bibfnamefont {V.}~\bibnamefont {Baibhav}},\
  and\ \bibinfo {author} {\bibfnamefont {R.}~\bibnamefont {Cotesta}},\
  }\bibfield  {title} {\bibinfo {title} {{Extracting linear and nonlinear
  quasinormal modes from black hole merger simulations}},\ }\href
  {https://doi.org/10.1103/PhysRevD.109.044069} {\bibfield  {journal} {\bibinfo
   {journal} {Phys. Rev. D}\ }\textbf {\bibinfo {volume} {109}},\ \bibinfo
  {pages} {044069} (\bibinfo {year} {2024})},\ \bibinfo {note} {[Erratum:
  Phys.Rev.D 110, 049902 (2024)]},\ \Eprint {https://arxiv.org/abs/2310.04489}
  {arXiv:2310.04489 [gr-qc]} \BibitemShut {NoStop}%
\bibitem [{\citenamefont {Ma}\ and\ \citenamefont
  {Yang}(2024)}]{ma2024excitation}%
  \BibitemOpen
  \bibfield  {author} {\bibinfo {author} {\bibfnamefont {S.}~\bibnamefont
  {Ma}}\ and\ \bibinfo {author} {\bibfnamefont {H.}~\bibnamefont {Yang}},\
  }\bibfield  {title} {\bibinfo {title} {{Excitation of quadratic quasinormal
  modes for Kerr black holes}},\ }\href
  {https://doi.org/10.1103/PhysRevD.109.104070} {\bibfield  {journal} {\bibinfo
   {journal} {Phys. Rev. D}\ }\textbf {\bibinfo {volume} {109}},\ \bibinfo
  {pages} {104070} (\bibinfo {year} {2024})},\ \Eprint
  {https://arxiv.org/abs/2401.15516} {arXiv:2401.15516 [gr-qc]} \BibitemShut
  {NoStop}%
\bibitem [{\citenamefont {Yi}\ \emph {et~al.}(2024)\citenamefont {Yi},
  \citenamefont {Kuntz}, \citenamefont {Barausse}, \citenamefont {Berti},
  \citenamefont {Cheung}, \citenamefont {Kritos},\ and\ \citenamefont
  {Maselli}}]{Yi:2024elj}%
  \BibitemOpen
  \bibfield  {author} {\bibinfo {author} {\bibfnamefont {S.}~\bibnamefont
  {Yi}}, \bibinfo {author} {\bibfnamefont {A.}~\bibnamefont {Kuntz}}, \bibinfo
  {author} {\bibfnamefont {E.}~\bibnamefont {Barausse}}, \bibinfo {author}
  {\bibfnamefont {E.}~\bibnamefont {Berti}}, \bibinfo {author} {\bibfnamefont
  {M.~H.-Y.}\ \bibnamefont {Cheung}}, \bibinfo {author} {\bibfnamefont
  {K.}~\bibnamefont {Kritos}},\ and\ \bibinfo {author} {\bibfnamefont
  {A.}~\bibnamefont {Maselli}},\ }\bibfield  {title} {\bibinfo {title}
  {{Nonlinear quasinormal mode detectability with next-generation gravitational
  wave detectors}},\ }\href {https://doi.org/10.1103/PhysRevD.109.124029}
  {\bibfield  {journal} {\bibinfo  {journal} {Phys. Rev. D}\ }\textbf {\bibinfo
  {volume} {109}},\ \bibinfo {pages} {124029} (\bibinfo {year} {2024})},\
  \Eprint {https://arxiv.org/abs/2403.09767} {arXiv:2403.09767 [gr-qc]}
  \BibitemShut {NoStop}%
\bibitem [{\citenamefont {Zhu}\ \emph {et~al.}(2024{\natexlab{b}})\citenamefont
  {Zhu} \emph {et~al.}}]{Zhu:2024rej}%
  \BibitemOpen
  \bibfield  {author} {\bibinfo {author} {\bibfnamefont {H.}~\bibnamefont
  {Zhu}} \emph {et~al.},\ }\bibfield  {title} {\bibinfo {title} {{Nonlinear
  effects in black hole ringdown from scattering experiments: Spin and initial
  data dependence of quadratic mode coupling}},\ }\href
  {https://doi.org/10.1103/PhysRevD.109.104050} {\bibfield  {journal} {\bibinfo
   {journal} {Phys. Rev. D}\ }\textbf {\bibinfo {volume} {109}},\ \bibinfo
  {pages} {104050} (\bibinfo {year} {2024}{\natexlab{b}})},\ \Eprint
  {https://arxiv.org/abs/2401.00805} {arXiv:2401.00805 [gr-qc]} \BibitemShut
  {NoStop}%
\bibitem [{\citenamefont {Bucciotti}\ \emph {et~al.}(2024)\citenamefont
  {Bucciotti}, \citenamefont {Juliano}, \citenamefont {Kuntz},\ and\
  \citenamefont {Trincherini}}]{Bucciotti:2024zyp}%
  \BibitemOpen
  \bibfield  {author} {\bibinfo {author} {\bibfnamefont {B.}~\bibnamefont
  {Bucciotti}}, \bibinfo {author} {\bibfnamefont {L.}~\bibnamefont {Juliano}},
  \bibinfo {author} {\bibfnamefont {A.}~\bibnamefont {Kuntz}},\ and\ \bibinfo
  {author} {\bibfnamefont {E.}~\bibnamefont {Trincherini}},\ }\bibfield
  {title} {\bibinfo {title} {{Quadratic quasinormal modes of a Schwarzschild
  black hole}},\ }\href {https://doi.org/10.1103/PhysRevD.110.104048}
  {\bibfield  {journal} {\bibinfo  {journal} {Phys. Rev. D}\ }\textbf {\bibinfo
  {volume} {110}},\ \bibinfo {pages} {104048} (\bibinfo {year} {2024})},\
  \Eprint {https://arxiv.org/abs/2405.06012} {arXiv:2405.06012 [gr-qc]}
  \BibitemShut {NoStop}%
\bibitem [{\citenamefont {Panosso~Macedo}\ and\ \citenamefont
  {Ansorg}(2014)}]{PanossoMacedo:2014dnr}%
  \BibitemOpen
  \bibfield  {author} {\bibinfo {author} {\bibfnamefont {R.}~\bibnamefont
  {Panosso~Macedo}}\ and\ \bibinfo {author} {\bibfnamefont {M.}~\bibnamefont
  {Ansorg}},\ }\bibfield  {title} {\bibinfo {title} {{Axisymmetric fully
  spectral code for hyperbolic equations}},\ }\href
  {https://doi.org/10.1016/j.jcp.2014.07.040} {\bibfield  {journal} {\bibinfo
  {journal} {J. Comput. Phys.}\ }\textbf {\bibinfo {volume} {276}},\ \bibinfo
  {pages} {357} (\bibinfo {year} {2014})},\ \Eprint
  {https://arxiv.org/abs/1402.7343} {arXiv:1402.7343 [physics.comp-ph]}
  \BibitemShut {NoStop}%
\bibitem [{\citenamefont {Ansorg}\ and\ \citenamefont
  {Panosso~Macedo}(2016)}]{Ansorg:2016ztf}%
  \BibitemOpen
  \bibfield  {author} {\bibinfo {author} {\bibfnamefont {M.}~\bibnamefont
  {Ansorg}}\ and\ \bibinfo {author} {\bibfnamefont {R.}~\bibnamefont
  {Panosso~Macedo}},\ }\bibfield  {title} {\bibinfo {title} {{Spectral
  decomposition of black-hole perturbations on hyperboloidal slices}},\ }\href
  {https://doi.org/10.1103/PhysRevD.93.124016} {\bibfield  {journal} {\bibinfo
  {journal} {Phys. Rev. D}\ }\textbf {\bibinfo {volume} {93}},\ \bibinfo
  {pages} {124016} (\bibinfo {year} {2016})},\ \Eprint
  {https://arxiv.org/abs/1604.02261} {arXiv:1604.02261 [gr-qc]} \BibitemShut
  {NoStop}%
\bibitem [{\citenamefont {Ammon}\ \emph {et~al.}(2016)\citenamefont {Ammon},
  \citenamefont {Grieninger}, \citenamefont {Jimenez-Alba}, \citenamefont
  {Macedo},\ and\ \citenamefont {Melgar}}]{Ammon:2016fru}%
  \BibitemOpen
  \bibfield  {author} {\bibinfo {author} {\bibfnamefont {M.}~\bibnamefont
  {Ammon}}, \bibinfo {author} {\bibfnamefont {S.}~\bibnamefont {Grieninger}},
  \bibinfo {author} {\bibfnamefont {A.}~\bibnamefont {Jimenez-Alba}}, \bibinfo
  {author} {\bibfnamefont {R.~P.}\ \bibnamefont {Macedo}},\ and\ \bibinfo
  {author} {\bibfnamefont {L.}~\bibnamefont {Melgar}},\ }\bibfield  {title}
  {\bibinfo {title} {{Holographic quenches and anomalous transport}},\ }\href
  {https://doi.org/10.1007/JHEP09(2016)131} {\bibfield  {journal} {\bibinfo
  {journal} {JHEP}\ }\textbf {\bibinfo {volume} {09}},\ \bibinfo {pages}
  {131}},\ \Eprint {https://arxiv.org/abs/1607.06817} {arXiv:1607.06817
  [hep-th]} \BibitemShut {NoStop}%
\bibitem [{\citenamefont {Panosso~Macedo}\ \emph {et~al.}(2018)\citenamefont
  {Panosso~Macedo}, \citenamefont {Jaramillo},\ and\ \citenamefont
  {Ansorg}}]{PanossoMacedo:2018hab}%
  \BibitemOpen
  \bibfield  {author} {\bibinfo {author} {\bibfnamefont {R.}~\bibnamefont
  {Panosso~Macedo}}, \bibinfo {author} {\bibfnamefont {J.~L.}\ \bibnamefont
  {Jaramillo}},\ and\ \bibinfo {author} {\bibfnamefont {M.}~\bibnamefont
  {Ansorg}},\ }\bibfield  {title} {\bibinfo {title} {{Hyperboloidal slicing
  approach to quasi-normal mode expansions: the Reissner-Nordstr\"om case}},\
  }\href {https://doi.org/10.1103/PhysRevD.98.124005} {\bibfield  {journal}
  {\bibinfo  {journal} {Phys. Rev. D}\ }\textbf {\bibinfo {volume} {98}},\
  \bibinfo {pages} {124005} (\bibinfo {year} {2018})},\ \Eprint
  {https://arxiv.org/abs/1809.02837} {arXiv:1809.02837 [gr-qc]} \BibitemShut
  {NoStop}%
\bibitem [{\citenamefont {Panosso~Macedo}(2020)}]{PanossoMacedo:2019npm}%
  \BibitemOpen
  \bibfield  {author} {\bibinfo {author} {\bibfnamefont {R.}~\bibnamefont
  {Panosso~Macedo}},\ }\bibfield  {title} {\bibinfo {title} {{Hyperboloidal
  framework for the Kerr spacetime}},\ }\href
  {https://doi.org/10.1088/1361-6382/ab6e3e} {\bibfield  {journal} {\bibinfo
  {journal} {Class. Quant. Grav.}\ }\textbf {\bibinfo {volume} {37}},\ \bibinfo
  {pages} {065019} (\bibinfo {year} {2020})},\ \Eprint
  {https://arxiv.org/abs/1910.13452} {arXiv:1910.13452 [gr-qc]} \BibitemShut
  {NoStop}%
\bibitem [{\citenamefont {Jaramillo}\ \emph {et~al.}(2021)\citenamefont
  {Jaramillo}, \citenamefont {Panosso~Macedo},\ and\ \citenamefont
  {Al~Sheikh}}]{Jaramillo:2020tuu}%
  \BibitemOpen
  \bibfield  {author} {\bibinfo {author} {\bibfnamefont {J.~L.}\ \bibnamefont
  {Jaramillo}}, \bibinfo {author} {\bibfnamefont {R.}~\bibnamefont
  {Panosso~Macedo}},\ and\ \bibinfo {author} {\bibfnamefont {L.}~\bibnamefont
  {Al~Sheikh}},\ }\bibfield  {title} {\bibinfo {title} {{Pseudospectrum and
  Black Hole Quasinormal Mode Instability}},\ }\href
  {https://doi.org/10.1103/PhysRevX.11.031003} {\bibfield  {journal} {\bibinfo
  {journal} {Phys. Rev. X}\ }\textbf {\bibinfo {volume} {11}},\ \bibinfo
  {pages} {031003} (\bibinfo {year} {2021})},\ \Eprint
  {https://arxiv.org/abs/2004.06434} {arXiv:2004.06434 [gr-qc]} \BibitemShut
  {NoStop}%
\bibitem [{\citenamefont {Panosso~Macedo}(2024)}]{PanossoMacedo:2023qzp}%
  \BibitemOpen
  \bibfield  {author} {\bibinfo {author} {\bibfnamefont {R.}~\bibnamefont
  {Panosso~Macedo}},\ }\bibfield  {title} {\bibinfo {title} {{Hyperboloidal
  approach for static spherically symmetric spacetimes: a didactical
  introduction and applications in black-hole physics}},\ }\href
  {https://doi.org/10.1098/rsta.2023.0046} {\bibfield  {journal} {\bibinfo
  {journal} {Phil. Trans. Roy. Soc. Lond. A}\ }\textbf {\bibinfo {volume}
  {382}},\ \bibinfo {pages} {20230046} (\bibinfo {year} {2024})},\ \Eprint
  {https://arxiv.org/abs/2307.15735} {arXiv:2307.15735 [gr-qc]} \BibitemShut
  {NoStop}%
\bibitem [{\citenamefont {Panosso~Macedo}\ \emph {et~al.}(2022)\citenamefont
  {Panosso~Macedo}, \citenamefont {Leather}, \citenamefont {Warburton},
  \citenamefont {Wardell},\ and\ \citenamefont
  {Zengino\u{g}lu}}]{PanossoMacedo:2022fdi}%
  \BibitemOpen
  \bibfield  {author} {\bibinfo {author} {\bibfnamefont {R.}~\bibnamefont
  {Panosso~Macedo}}, \bibinfo {author} {\bibfnamefont {B.}~\bibnamefont
  {Leather}}, \bibinfo {author} {\bibfnamefont {N.}~\bibnamefont {Warburton}},
  \bibinfo {author} {\bibfnamefont {B.}~\bibnamefont {Wardell}},\ and\ \bibinfo
  {author} {\bibfnamefont {A.}~\bibnamefont {Zengino\u{g}lu}},\ }\bibfield
  {title} {\bibinfo {title} {{Hyperboloidal method for frequency-domain
  self-force calculations}},\ }\href
  {https://doi.org/10.1103/PhysRevD.105.104033} {\bibfield  {journal} {\bibinfo
   {journal} {Phys. Rev. D}\ }\textbf {\bibinfo {volume} {105}},\ \bibinfo
  {pages} {104033} (\bibinfo {year} {2022})},\ \Eprint
  {https://arxiv.org/abs/2202.01794} {arXiv:2202.01794 [gr-qc]} \BibitemShut
  {NoStop}%
\bibitem [{\citenamefont {Spiers}\ \emph
  {et~al.}(2024{\natexlab{a}})\citenamefont {Spiers}, \citenamefont {Pound},\
  and\ \citenamefont {Wardell}}]{Spiers:2023mor}%
  \BibitemOpen
  \bibfield  {author} {\bibinfo {author} {\bibfnamefont {A.}~\bibnamefont
  {Spiers}}, \bibinfo {author} {\bibfnamefont {A.}~\bibnamefont {Pound}},\ and\
  \bibinfo {author} {\bibfnamefont {B.}~\bibnamefont {Wardell}},\ }\bibfield
  {title} {\bibinfo {title} {{Second-order perturbations of the Schwarzschild
  spacetime: Practical, covariant, and gauge-invariant formalisms}},\ }\href
  {https://doi.org/10.1103/PhysRevD.110.064030} {\bibfield  {journal} {\bibinfo
   {journal} {Phys. Rev. D}\ }\textbf {\bibinfo {volume} {110}},\ \bibinfo
  {pages} {064030} (\bibinfo {year} {2024}{\natexlab{a}})},\ \Eprint
  {https://arxiv.org/abs/2306.17847} {arXiv:2306.17847 [gr-qc]} \BibitemShut
  {NoStop}%
\bibitem [{\citenamefont {Spiers}\ \emph {et~al.}(2023)\citenamefont {Spiers},
  \citenamefont {Pound},\ and\ \citenamefont {Moxon}}]{mySecond-orderTeuk}%
  \BibitemOpen
  \bibfield  {author} {\bibinfo {author} {\bibfnamefont {A.}~\bibnamefont
  {Spiers}}, \bibinfo {author} {\bibfnamefont {A.}~\bibnamefont {Pound}},\ and\
  \bibinfo {author} {\bibfnamefont {J.}~\bibnamefont {Moxon}},\ }\bibfield
  {title} {\bibinfo {title} {{Second-order Teukolsky formalism in Kerr
  spacetime: Formulation and nonlinear source}},\ }\href
  {https://doi.org/10.1103/PhysRevD.108.064002} {\bibfield  {journal} {\bibinfo
   {journal} {Phys. Rev. D}\ }\textbf {\bibinfo {volume} {108}},\ \bibinfo
  {pages} {064002} (\bibinfo {year} {2023})},\ \Eprint
  {https://arxiv.org/abs/2305.19332} {arXiv:2305.19332 [gr-qc]} \BibitemShut
  {NoStop}%
\bibitem [{\citenamefont {Pound}\ and\ \citenamefont
  {Wardell}(2021)}]{Pound:2021qin}%
  \BibitemOpen
  \bibfield  {author} {\bibinfo {author} {\bibfnamefont {A.}~\bibnamefont
  {Pound}}\ and\ \bibinfo {author} {\bibfnamefont {B.}~\bibnamefont
  {Wardell}},\ }\bibfield  {title} {\bibinfo {title} {{Black hole perturbation
  theory and gravitational self-force}}\ }\href
  {https://doi.org/10.1007/978-981-15-4702-7\_38-1}
  {10.1007/978-981-15-4702-7\_38-1} (\bibinfo {year} {2021}),\ \Eprint
  {https://arxiv.org/abs/2101.04592} {arXiv:2101.04592 [gr-qc]} \BibitemShut
  {NoStop}%
\bibitem [{\citenamefont {Teukolsky}(1972)}]{Teukolsky:1972my}%
  \BibitemOpen
  \bibfield  {author} {\bibinfo {author} {\bibfnamefont {S.~A.}\ \bibnamefont
  {Teukolsky}},\ }\bibfield  {title} {\bibinfo {title} {{Rotating black holes -
  separable wave equations for gravitational and electromagnetic
  perturbations}},\ }\href {https://doi.org/10.1103/PhysRevLett.29.1114}
  {\bibfield  {journal} {\bibinfo  {journal} {Phys. Rev. Lett.}\ }\textbf
  {\bibinfo {volume} {29}},\ \bibinfo {pages} {1114} (\bibinfo {year}
  {1972})}\BibitemShut {NoStop}%
\bibitem [{\citenamefont {Teukolsky}(1973)}]{Teukolsky:1973ha}%
  \BibitemOpen
  \bibfield  {author} {\bibinfo {author} {\bibfnamefont {S.~A.}\ \bibnamefont
  {Teukolsky}},\ }\bibfield  {title} {\bibinfo {title} {{Perturbations of a
  rotating black hole. 1. Fundamental equations for gravitational
  electromagnetic and neutrino field perturbations}},\ }\href
  {https://doi.org/10.1086/152444} {\bibfield  {journal} {\bibinfo  {journal}
  {Astrophys. J.}\ }\textbf {\bibinfo {volume} {185}},\ \bibinfo {pages} {635}
  (\bibinfo {year} {1973})}\BibitemShut {NoStop}%
\bibitem [{\citenamefont {Leaver}(1986)}]{Leaver:1986gd}%
  \BibitemOpen
  \bibfield  {author} {\bibinfo {author} {\bibfnamefont {E.~W.}\ \bibnamefont
  {Leaver}},\ }\bibfield  {title} {\bibinfo {title} {{Spectral decomposition of
  the perturbation response of the Schwarzschild geometry}},\ }\href
  {https://doi.org/10.1103/PhysRevD.34.384} {\bibfield  {journal} {\bibinfo
  {journal} {Phys. Rev. D}\ }\textbf {\bibinfo {volume} {34}},\ \bibinfo
  {pages} {384} (\bibinfo {year} {1986})}\BibitemShut {NoStop}%
\bibitem [{\citenamefont {Zenginoglu}(2011)}]{Zenginoglu:2011jz}%
  \BibitemOpen
  \bibfield  {author} {\bibinfo {author} {\bibfnamefont {A.}~\bibnamefont
  {Zenginoglu}},\ }\bibfield  {title} {\bibinfo {title} {{A Geometric framework
  for black hole perturbations}},\ }\href
  {https://doi.org/10.1103/PhysRevD.83.127502} {\bibfield  {journal} {\bibinfo
  {journal} {Phys. Rev. D}\ }\textbf {\bibinfo {volume} {83}},\ \bibinfo
  {pages} {127502} (\bibinfo {year} {2011})},\ \Eprint
  {https://arxiv.org/abs/1102.2451} {arXiv:1102.2451 [gr-qc]} \BibitemShut
  {NoStop}%
\bibitem [{\citenamefont {Dhani}(2021)}]{Dhani:2020nik}%
  \BibitemOpen
  \bibfield  {author} {\bibinfo {author} {\bibfnamefont {A.}~\bibnamefont
  {Dhani}},\ }\bibfield  {title} {\bibinfo {title} {{Importance of mirror modes
  in binary black hole ringdown waveform}},\ }\href
  {https://doi.org/10.1103/PhysRevD.103.104048} {\bibfield  {journal} {\bibinfo
   {journal} {Phys. Rev. D}\ }\textbf {\bibinfo {volume} {103}},\ \bibinfo
  {pages} {104048} (\bibinfo {year} {2021})},\ \Eprint
  {https://arxiv.org/abs/2010.08602} {arXiv:2010.08602 [gr-qc]} \BibitemShut
  {NoStop}%
\bibitem [{\citenamefont {M\"adler}\ and\ \citenamefont
  {Winicour}(2016)}]{Madler:2016xju}%
  \BibitemOpen
  \bibfield  {author} {\bibinfo {author} {\bibfnamefont {T.}~\bibnamefont
  {M\"adler}}\ and\ \bibinfo {author} {\bibfnamefont {J.}~\bibnamefont
  {Winicour}},\ }\bibfield  {title} {\bibinfo {title} {{Bondi-Sachs
  Formalism}},\ }\href {https://doi.org/10.4249/scholarpedia.33528} {\bibfield
  {journal} {\bibinfo  {journal} {Scholarpedia}\ }\textbf {\bibinfo {volume}
  {11}},\ \bibinfo {pages} {33528} (\bibinfo {year} {2016})},\ \Eprint
  {https://arxiv.org/abs/1609.01731} {arXiv:1609.01731 [gr-qc]} \BibitemShut
  {NoStop}%
\bibitem [{\citenamefont {Martel}\ and\ \citenamefont
  {Poisson}(2005)}]{Martel:2005ir}%
  \BibitemOpen
  \bibfield  {author} {\bibinfo {author} {\bibfnamefont {K.}~\bibnamefont
  {Martel}}\ and\ \bibinfo {author} {\bibfnamefont {E.}~\bibnamefont
  {Poisson}},\ }\bibfield  {title} {\bibinfo {title} {{Gravitational
  perturbations of the Schwarzschild spacetime: A Practical covariant and
  gauge-invariant formalism}},\ }\href
  {https://doi.org/10.1103/PhysRevD.71.104003} {\bibfield  {journal} {\bibinfo
  {journal} {Phys. Rev. D}\ }\textbf {\bibinfo {volume} {71}},\ \bibinfo
  {pages} {104003} (\bibinfo {year} {2005})},\ \Eprint
  {https://arxiv.org/abs/gr-qc/0502028} {arXiv:gr-qc/0502028} \BibitemShut
  {NoStop}%
\bibitem [{\citenamefont {Campanelli}\ and\ \citenamefont
  {Lousto}(1999)}]{Campanelli:1998jv}%
  \BibitemOpen
  \bibfield  {author} {\bibinfo {author} {\bibfnamefont {M.}~\bibnamefont
  {Campanelli}}\ and\ \bibinfo {author} {\bibfnamefont {C.~O.}\ \bibnamefont
  {Lousto}},\ }\bibfield  {title} {\bibinfo {title} {{Second order gauge
  invariant gravitational perturbations of a Kerr black hole}},\ }\href
  {https://doi.org/10.1103/PhysRevD.59.124022} {\bibfield  {journal} {\bibinfo
  {journal} {Phys. Rev. D}\ }\textbf {\bibinfo {volume} {59}},\ \bibinfo
  {pages} {124022} (\bibinfo {year} {1999})},\ \Eprint
  {https://arxiv.org/abs/gr-qc/9811019} {arXiv:gr-qc/9811019} \BibitemShut
  {NoStop}%
\bibitem [{\citenamefont {Green}\ \emph {et~al.}(2020)\citenamefont {Green},
  \citenamefont {Hollands},\ and\ \citenamefont {Zimmerman}}]{Green:2019nam}%
  \BibitemOpen
  \bibfield  {author} {\bibinfo {author} {\bibfnamefont {S.~R.}\ \bibnamefont
  {Green}}, \bibinfo {author} {\bibfnamefont {S.}~\bibnamefont {Hollands}},\
  and\ \bibinfo {author} {\bibfnamefont {P.}~\bibnamefont {Zimmerman}},\
  }\bibfield  {title} {\bibinfo {title} {{Teukolsky formalism for nonlinear
  Kerr perturbations}},\ }\href {https://doi.org/10.1088/1361-6382/ab7075}
  {\bibfield  {journal} {\bibinfo  {journal} {Class. Quant. Grav.}\ }\textbf
  {\bibinfo {volume} {37}},\ \bibinfo {pages} {075001} (\bibinfo {year}
  {2020})},\ \Eprint {https://arxiv.org/abs/1908.09095} {arXiv:1908.09095
  [gr-qc]} \BibitemShut {NoStop}%
\bibitem [{\citenamefont {Spiers}\ \emph
  {et~al.}(2024{\natexlab{b}})\citenamefont {Spiers}, \citenamefont {Wardell},
  \citenamefont {Pound}, \citenamefont {Upton},\ and\ \citenamefont
  {Warburton}}]{warburton2023perturbationequations}%
  \BibitemOpen
  \bibfield  {author} {\bibinfo {author} {\bibfnamefont {A.}~\bibnamefont
  {Spiers}}, \bibinfo {author} {\bibfnamefont {B.}~\bibnamefont {Wardell}},
  \bibinfo {author} {\bibfnamefont {A.}~\bibnamefont {Pound}}, \bibinfo
  {author} {\bibfnamefont {S.~D.}\ \bibnamefont {Upton}},\ and\ \bibinfo
  {author} {\bibfnamefont {N.}~\bibnamefont {Warburton}},\ }\href
  {https://doi.org/10.5281/zenodo.11199718} {\bibinfo {title}
  {Perturbation{E}quations}} (\bibinfo {year} {2024}{\natexlab{b}})\BibitemShut
  {NoStop}%
\bibitem [{\citenamefont {Okuzumi}\ \emph {et~al.}(2008)\citenamefont
  {Okuzumi}, \citenamefont {Ioka},\ and\ \citenamefont
  {Sakagami}}]{Okuzumi:2008ej}%
  \BibitemOpen
  \bibfield  {author} {\bibinfo {author} {\bibfnamefont {S.}~\bibnamefont
  {Okuzumi}}, \bibinfo {author} {\bibfnamefont {K.}~\bibnamefont {Ioka}},\ and\
  \bibinfo {author} {\bibfnamefont {M.}~\bibnamefont {Sakagami}},\ }\bibfield
  {title} {\bibinfo {title} {{Possible Discovery of Nonlinear Tail and
  Quasinormal Modes in Black Hole Ringdown}},\ }\href
  {https://doi.org/10.1103/PhysRevD.77.124018} {\bibfield  {journal} {\bibinfo
  {journal} {Phys. Rev. D}\ }\textbf {\bibinfo {volume} {77}},\ \bibinfo
  {pages} {124018} (\bibinfo {year} {2008})},\ \Eprint
  {https://arxiv.org/abs/0803.0501} {arXiv:0803.0501 [gr-qc]} \BibitemShut
  {NoStop}%
\bibitem [{\citenamefont {Lagos}\ and\ \citenamefont
  {Hui}(2023)}]{Lagos:2022otp}%
  \BibitemOpen
  \bibfield  {author} {\bibinfo {author} {\bibfnamefont {M.}~\bibnamefont
  {Lagos}}\ and\ \bibinfo {author} {\bibfnamefont {L.}~\bibnamefont {Hui}},\
  }\bibfield  {title} {\bibinfo {title} {{Generation and propagation of
  nonlinear quasinormal modes of a Schwarzschild black hole}},\ }\href
  {https://doi.org/10.1103/PhysRevD.107.044040} {\bibfield  {journal} {\bibinfo
   {journal} {Phys. Rev. D}\ }\textbf {\bibinfo {volume} {107}},\ \bibinfo
  {pages} {044040} (\bibinfo {year} {2023})},\ \Eprint
  {https://arxiv.org/abs/2208.07379} {arXiv:2208.07379 [gr-qc]} \BibitemShut
  {NoStop}%
\bibitem [{\citenamefont {Spiers}(2023)}]{2nd-order-notebook}%
  \BibitemOpen
  \bibfield  {author} {\bibinfo {author} {\bibfnamefont {A.}~\bibnamefont
  {Spiers}},\ }\href@noop {} {\bibinfo {title} {{NP and GHP Formalisms for
  second-order Teukolsky equations}}},\ \bibinfo {howpublished}
  {\url{https://github.com/DrAndrewSpiers/NP-and-GHP-Formalisms-for-2nd-order-Teukolsky}}
  (\bibinfo {year} {2023})\BibitemShut {NoStop}%
\bibitem [{\citenamefont {Perrone}\ \emph {et~al.}(2024)\citenamefont
  {Perrone}, \citenamefont {Barreira}, \citenamefont {Kehagias},\ and\
  \citenamefont {Riotto}}]{Perrone:2023jzq}%
  \BibitemOpen
  \bibfield  {author} {\bibinfo {author} {\bibfnamefont {D.}~\bibnamefont
  {Perrone}}, \bibinfo {author} {\bibfnamefont {T.}~\bibnamefont {Barreira}},
  \bibinfo {author} {\bibfnamefont {A.}~\bibnamefont {Kehagias}},\ and\
  \bibinfo {author} {\bibfnamefont {A.}~\bibnamefont {Riotto}},\ }\bibfield
  {title} {\bibinfo {title} {{Non-linear black hole ringdowns: An analytical
  approach}},\ }\href {https://doi.org/10.1016/j.nuclphysb.2023.116432}
  {\bibfield  {journal} {\bibinfo  {journal} {Nucl. Phys. B}\ }\textbf
  {\bibinfo {volume} {999}},\ \bibinfo {pages} {116432} (\bibinfo {year}
  {2024})},\ \Eprint {https://arxiv.org/abs/2308.15886} {arXiv:2308.15886
  [gr-qc]} \BibitemShut {NoStop}%
\bibitem [{\citenamefont {Nakano}\ and\ \citenamefont
  {Ioka}(2007)}]{Nakano:2007cj}%
  \BibitemOpen
  \bibfield  {author} {\bibinfo {author} {\bibfnamefont {H.}~\bibnamefont
  {Nakano}}\ and\ \bibinfo {author} {\bibfnamefont {K.}~\bibnamefont {Ioka}},\
  }\bibfield  {title} {\bibinfo {title} {{Second Order Quasi-Normal Mode of the
  Schwarzschild Black Hole}},\ }\href
  {https://doi.org/10.1103/PhysRevD.76.084007} {\bibfield  {journal} {\bibinfo
  {journal} {Phys. Rev. D}\ }\textbf {\bibinfo {volume} {76}},\ \bibinfo
  {pages} {084007} (\bibinfo {year} {2007})},\ \Eprint
  {https://arxiv.org/abs/0708.0450} {arXiv:0708.0450 [gr-qc]} \BibitemShut
  {NoStop}%
\bibitem [{\citenamefont {Spiers}(2024)}]{Spiers:2024src}%
  \BibitemOpen
  \bibfield  {author} {\bibinfo {author} {\bibfnamefont {A.}~\bibnamefont
  {Spiers}},\ }\bibfield  {title} {\bibinfo {title} {{Analytically separating
  the source of the Teukolsky equation}},\ }\href
  {https://doi.org/10.1103/PhysRevD.109.104059} {\bibfield  {journal} {\bibinfo
   {journal} {Phys. Rev. D}\ }\textbf {\bibinfo {volume} {109}},\ \bibinfo
  {pages} {104059} (\bibinfo {year} {2024})},\ \Eprint
  {https://arxiv.org/abs/2402.00604} {arXiv:2402.00604 [gr-qc]} \BibitemShut
  {NoStop}%
\bibitem [{\citenamefont {Mitman}\ \emph {et~al.}(2024)\citenamefont {Mitman}
  \emph {et~al.}}]{Mitman:2024uss}%
  \BibitemOpen
  \bibfield  {author} {\bibinfo {author} {\bibfnamefont {K.}~\bibnamefont
  {Mitman}} \emph {et~al.},\ }\bibfield  {title} {\bibinfo {title} {{A Review
  of Gravitational Memory and BMS Frame Fixing in Numerical Relativity}},\
  }\href@noop {} {\  (\bibinfo {year} {2024})},\ \Eprint
  {https://arxiv.org/abs/2405.08868} {arXiv:2405.08868 [gr-qc]} \BibitemShut
  {NoStop}%
\bibitem [{\citenamefont {Cardoso}\ \emph {et~al.}(2024)\citenamefont
  {Cardoso}, \citenamefont {Carullo}, \citenamefont {De~Amicis}, \citenamefont
  {Duque}, \citenamefont {Katagiri}, \citenamefont {Pereniguez}, \citenamefont
  {Redondo-Yuste}, \citenamefont {Spieksma},\ and\ \citenamefont
  {Zhong}}]{Cardoso:2024jme}%
  \BibitemOpen
  \bibfield  {author} {\bibinfo {author} {\bibfnamefont {V.}~\bibnamefont
  {Cardoso}}, \bibinfo {author} {\bibfnamefont {G.}~\bibnamefont {Carullo}},
  \bibinfo {author} {\bibfnamefont {M.}~\bibnamefont {De~Amicis}}, \bibinfo
  {author} {\bibfnamefont {F.}~\bibnamefont {Duque}}, \bibinfo {author}
  {\bibfnamefont {T.}~\bibnamefont {Katagiri}}, \bibinfo {author}
  {\bibfnamefont {D.}~\bibnamefont {Pereniguez}}, \bibinfo {author}
  {\bibfnamefont {J.}~\bibnamefont {Redondo-Yuste}}, \bibinfo {author}
  {\bibfnamefont {T.~F.~M.}\ \bibnamefont {Spieksma}},\ and\ \bibinfo {author}
  {\bibfnamefont {Z.}~\bibnamefont {Zhong}},\ }\bibfield  {title} {\bibinfo
  {title} {{Hushing black holes: Tails in dynamical spacetimes}},\ }\href
  {https://doi.org/10.1103/PhysRevD.109.L121502} {\bibfield  {journal}
  {\bibinfo  {journal} {Phys. Rev. D}\ }\textbf {\bibinfo {volume} {109}},\
  \bibinfo {pages} {L121502} (\bibinfo {year} {2024})},\ \Eprint
  {https://arxiv.org/abs/2405.12290} {arXiv:2405.12290 [gr-qc]} \BibitemShut
  {NoStop}%
\end{thebibliography}%

\clearpage

%%%%%%%%%%%%%%%%%%%%%%%%%%%%%%%%%%%%%%%%%%%%%%%%%%
%%%%%%%%%%%%%%%%% Supplemental Material %%%%%%%%%%%%%%%%%%%%%
%%%%%%%%%%%%%%%%%%%%%%%%%%%%%%%%%%%%%%%%%%%%%%%%%%

\section*{Supplemental Material}

In Table~\ref{table:sab}, we give the values for the frequencies $s_{\l\m\n}$ and the coefficients appearing in Eqs.~\eqref{eqn:Psi42ndOrder++}--\eqref{eqn:Psi42ndOrder--}, evaluated at $\scri^+$ ($\sigma=0$), for different $\l\m$ modes.
Note that for $\m=0$, Eqs \eqref{eqn:Psi42ndOrder++}-\eqref{eqn:Psi42ndOrder--} still remain valid, where the two amplitudes $(A_{\l\m\n}, A_{\l-\m\n}) \to (A_{\l 0 \n,+}, A_{\l 0 \n,-})$ are associated with the frequencies, $s_{\l 0\n, +}$ and $s_{\l 0\n, -}$ respectively.
\begin{table}[h] \centering
\begin{tabular}{ l|c c} 
\toprule
$(\l,\m,\n)$ & \multicolumn{2}{c}{at $\scri^+ \, (\sigma=0)$} \\
\midrule
\multirow{5}{4em}{$(2,0,0)$}
& $s_{200,+}$ & $ -0.0889623 - 0.373672 i $ \\ 
& $a^{40}$ & $ 0.617742 + 0.249534 i $ \\ 
& $b^{40}$ & $ 0.0695369 + 0.0405164 i $ \\ 
& $c^{40}$ & $ (0.0820959 - 5.27328 i) \times 10^{-4} $ \\ 
& $d^{40}$ & $ 2.26172 \times 10^{-4} $ \\
\midrule
\multirow{5}{4em}{$(2,2,0)$}
& $s_{220}$ & $ -0.0889623 - 0.373672 i $ \\ 
& $a^{44}$ & $ 0.861400 + 0.347958 i $ \\ 
& $b^{44}$ & $ 0.096965 + 0.056497 i $ \\ 
& $c^{40}$ & $ (0.136826 - 8.78887 i) \times 10^{-5} $ \\ 
& $d^{40}$ & $ 3.76953 \times 10^{-5} $ \\
\midrule
\multirow{5}{4em}{$(3,3,0)$}
& $s_{330}$ & $ -0.092703 - 0.599443 i $ \\ 
& $a^{66}$ & $ 0.55259 + 0.0645167 i $ \\ 
& $b^{66}$ & $ -0.0680453 - 0.0370004 i $ \\ 
& $c^{60}$ & $ (-7.59687 + 8.14475 i) \times 10^{-6} $ \\ 
& $d^{60}$ & $ 4.27944 \times 10^{-7} $ \\
\midrule
\multirow{5}{4em}{$(4,4,0)$}
& $s_{440}$ & $ -0.094164 - 0.809178 i $ \\ 
& $a^{88}$ & $ 0.318602 - 0.00475926 i $ \\ 
& $b^{88}$ & $ 0.0410101 + 0.0194345 i $ \\ 
& $c^{80}$ & $ (1.13628 - 0.744193 i) \times 10^{-6} $ \\ 
& $d^{80}$ & $ 1.20040 \times 10^{-8} $ \\
\bottomrule
\end{tabular}
\caption{Values of $s_{\l\m\n}$ and of the coefficients at $\scri^+$ appearing in Eqs~\eqref{eqn:Psi42ndOrder++}--\eqref{eqn:Psi42ndOrder--}, for a selection of mode numbers $(\l,\m,\n$).}
\label{table:sab}
\end{table}

For completeness, we give in Table~\ref{table:QQNM_Literature} the QQNM ratios as reported in the literature that are used to generate the data points in Fig.~\ref{fig:qqnm_ratio_440}.
These data points were computed in the following manner:
\begin{table} \centering
\begin{tabular}{ l c} 
\toprule
References & $\left(\mathcal{R}^{h_{\rm{SpEC}}} \right)^{44}$ \\
\midrule
Ma \emph{et al.} \cite{ma2024excitation} & $ 0.137 e^{-0.083 i} $ \\
Bucciotti \emph{et al} \cite{Bucciotti:2024zyp} & $ 0.154 e^{-0.068 i} $ \\
Cheung (1) \emph{et al.} \cite{Cheung:2022rbm} & $ 0.1637 e^{0.177 i} $ \\ 
Cheung (2) \emph{et al.} \cite{Cheung:2023vki} & $ 0.170 e^{0.077 i} $ \\ 
Zhu \emph{et al} \cite{Zhu:2024rej} & (by eye) $ 0.17 e^{-0.083 i} $ \\
Mitman \emph{et al} \cite{Mitman:2022qdl} & $ (0.15:0.20) e^{(-0.4:0.4) i} $ \\
Redondo-Yuste \emph{et al} \cite{Redondo-Yuste:2023seq} & $ 0.164 e^{-0.065 i} $ \\
\bottomrule
\end{tabular}
\caption{The reported values for the magnitude and phase of the QQNM ratio $\left(\mathcal{R}^{h_{\rm{SpEC}}} \right)^{44}$.}
\label{table:QQNM_Literature}
\end{table}

To include the data from Table~\ref{table:QQNM_Literature} in Fig.~\ref{fig:qqnm_ratio_440}, we must extract the ratio $C^-_{\l\m\n} / C^+_{\l\m\n}$ from the data. We do this starting from the values of the magnitude and phase of the QQNM ratio, $\left(\mathcal{R}^{h_{SpEC}} \right)^{LM}$ (as given in the cited references). Using relations \eqref{eqn:ratioSpEC}, \eqref{eqn:QQNMRatio}, and \eqref{eqn:Psi42ndOrder++}, together with Eqs.~\eqref{eqn:AtoC1} and \eqref{eqn:AtoC2}, we then calculate the (unique) complex ratio $C^-_{\l\m\n} / C^+_{\l\m\n}$. We can then plot the corresponding point in the complex plane, as shown in Fig.~\ref{fig:qqnm_ratio_440}. 

We emphasize that this map from $\left(\mathcal{R}^{h_{SpEC}} \right)^{LM}$ to $C^-_{\l\m\n} / C^+_{\l\m\n}$ depends on the coefficients in Table~\ref{table:sab}, which enter into Eq.~\eqref{eqn:Psi42ndOrder++}. Since the coefficients in Table~\ref{table:sab} assume a Schwarzschild background, our mapping yields an incorrect value for $C^-_{\l\m\n} / C^+_{\l\m\n}$ when applied to NR data for a spinning BH. However, the right panel of Fig.~\ref{fig:qqnm_ratio_440} displays NR results for nonspinning BHs, which are free from this error. The deviation of the NR data from the expected result $C^-_{20\n}=0$ in this case is broadly comparable to the deviation of the NR data in the left panel. This might suggest that the error in applying our mapping to spinning BHs can be comparable to the underlying error in the extraction of $\left(\mathcal{R}^{h_{SpEC}} \right)^{LM}$ from NR data.

The comparison to nonspinning NR data uses the QQNM ratio for the linear $(2,0,0)$ mode exciting the $(4,0)$ mode at second order. This is because the data is taken from simulations of head-on collisions, which only involve $\m=0$ modes due to axial symmetry. 
In this case, the reported values for the amplitude and phase, from Cheung (1), cannot be used as is, as their definition differ from ours. Since in a head-on collision only the even modes are excited, they combined the prograde and retrograde modes into a single ``real QNM'' and used a $\sin$ function (as opposed to a $\cos$ function) to fit the waveform. Accounting for this difference in conventions, we find their final result for the strain, associated with the frequency $2 s_{200,+}$, is
\begin{equation}
    (\mathcal{R}^{h_{SpEC}})^{40} = (0.125 \pm 0.0085) e^{i (-0.177 \pm 0.116)}.
\end{equation}

%%%%%%%%%%%%%%%%%%%%%%%%%%%%%%%%
%%%%%%%% End of Supplemental Material %%%%%%%%%
%%%%%%%%%%%%%%%%%%%%%%%%%%%%%%%%

\end{document}